\documentclass[12pt]{article}

\usepackage{newtxtext,newtxmath}

\usepackage{graphicx}

\usepackage[letterpaper,margin=1in]{geometry}

\renewenvironment{abstract}
	{\quotation}
	{\endquotation}

\date{}

\makeatletter
\renewcommand{\fnum@figure}{\textbf{Figure \thefigure}}
\renewcommand{\fnum@table}{\textbf{Table \thetable}}
\makeatother

\usepackage{scicite}

\usepackage{url}

\newcommand{\phispin}{$\Phi$SpIN}
\newcommand{\phispinsw}{$\Phi$SpIN-SW}

\def\scititle{
Physics-Spawned Neural Integration of Rotating Shallow-Water Dynamics}
\title{\bfseries \boldmath \scititle}

\author{
	Kaushal Gianchandani$^{1,2\ast}$,
    Luka Todorovic $^{3}$
    and
	Sai Ravela$^{2\dagger}$
    \and
	\small$^{1}$NASA Postdoctoral Program, NASA Goddard Institute for Space Studies, New York, NY, USA.\and
	\small$^{2}$Department of Earth, Atmospheric, and Planetary Science, Massachusetts Institute of Technology,\and \small Cambridge, MA, USA.\and  
    \small$^{3}$Department of Mathematics and Department of Statistics, University of Michigan, Ann Arbor, MI, USA \and
    \small$^{\ast,\dagger}$Corresponding authors. Emails: kaushalg@mit.edu ($\ast$), ravela@mit.edu ($\dagger$)
}

\begin{document}

\maketitle

\begin{abstract} 
Equation-derived neural integration can execute prescribed dynamics without training, while informative ensemble learning can estimate uncertain parameters and identify governing interactions without requiring differentiability. Together, these approaches motivate Physics-Spawned Informative Networks (\phispin), a framework for prescribed and adaptive hybrid dynamical systems. As a first step toward flood modeling applications, we present \phispinsw, its prescribed-physics realization for the rotating shallow-water equations. Compatible C-grid operators on spherical Voronoi meshes define a computational graph that retains non-polynomial operations and advances the state through recurrent neural integration compiled with JAX. Analytical benchmarks characterize numerical errors, and global and coastal experiments closely reproduce reference solutions generated using the Model Prediction Across Scales (MPAS) model. At approximately 14~km global resolution, A100, H200 and B200 GPUs attain 9-, 19- and 34-fold higher timestep throughput than a 96-core MPAS-SW CPU reference under matched workloads. These results extend training-free neural integration beyond polynomial dynamics to complex geophysical simulation, establishing a computational foundation for subsequent adjoint-free ensemble estimation of physical and neural parameters and informative adaptation of operator structure.
\end{abstract}

\section{Introduction}
An Earth system model is at once an executable statement of what is known and an approximation to what is not. Its governing equations specify resolved interactions, yet advancing them across finer grids, longer integrations and larger ensembles remains expensive \cite{kay2015community,fuhrer2018near}. At the same time, the equations used in a practical model are filtered, discretized and supplemented by scale-dependent approximations. Subgrid turbulence, transport and mixing remain unresolved, while constitutive laws, boundary interactions and forcing parameters may be uncertain. Improving such a model therefore requires two complementary capabilities: efficient execution of dynamics already prescribed by physical knowledge, and inference of the parameters and interactions that the resolved description cannot determine. The former should not have to be learned again in order to accomplish the latter.

Deep learning offers powerful alternatives to conventional simulation, from learned state-transition maps and neural operators to physics-informed solution approximations \cite{chen2018neural,li2021fourier,raissi2019physics,lam2023learning}. GraphCast demonstrates learned global forecasting \cite{lam2023learning}, while learned atmospheric parameterizations address processes that are not explicitly resolved \cite{rasp2018deep}. In hybrids such as NeuralGCM, a physical dynamical core evolves together with learned parameterizations trained through differentiable, end-to-end optimization \cite{kochkov2024neural}. These approaches show the value of data when a model is incomplete, but they also sharpen the question of what ought to be learned. If governing interactions and coefficients are already specified, fitting an emulator to reproduce them is not a prerequisite for executing them efficiently. GPU-native numerical models such as Oceananigans, Veros and ClimateMachine, as well as newer differentiable and JAX-based Earth-system implementations, already establish that prescribed physics can exploit accelerator computation directly \cite{ramadhan2020oceananigans,hafner2021fast,sridhar2022large,linck2026evaluating,yuan2026wrf,gentine2026legoesm}. The more consequential question is how prescribed and inferable dynamics can inhabit the same executable representation without forcing the model to relearn established interactions.

An early answer to the prescribed part of this question came from Trautner and Ravela's neural integration of continuous dynamics (2019) \cite{trautner2019}. They showed that polynomial dynamical equations could determine the topology and weights of polynomial networks (PolyNets), while compact recurrent circuits performed Runge--Kutta and Adams--Bashforth--Moulton time integration. The resulting network executed both the specified dynamical operator and its discrete-time evolution without training when the governing coefficients were known. It did not abolish numerical approximation: the spatial and temporal discretizations still determined the accuracy of the calculation. Rather, it removed the need to fit an additional approximation to dynamics already available in equation form. In this sense, neural integration is an execution principle, not a synonym for learning a surrogate.

The inference side of this problem has an earlier history in geophysical fluid dynamics. Ravela and McLaughlin developed fast ensemble smoothing methods for fixed-interval and fixed-lag estimation, showing how observations distributed through time could constrain model trajectories with modest additional cost relative to ensemble filtering and propagation \cite{ravela2007smoothing}. Ravela and colleagues subsequently coupled observations of a rotating laboratory fluid to the MIT general circulation model in real time, using ensemble-based assimilation to track the evolving fluid state \cite{ravela2010observatory}. These studies established practical links between prescribed fluid dynamics, observations and forward-ensemble inference, without requiring adjoint sensitivities. They estimated states rather than discovering governing terms or training neural-network weights, but provided an important precursor to treating the model itself as an object of ensemble inference.

The next step was therefore to infer uncertain dynamics, not only the state evolving under fixed dynamics. Universal Differential Equations (UDEs), introduced in 2020, coupled mechanistic equations with learned functions and demonstrated recovery of explicit relations through subsequent symbolic regression \cite{rackauckas2020ude}. Independently in 2020, Trautner, Margolis and Ravela introduced Informative Neural Ensemble Kalman Learning (IEKL), combining ensemble-based parameter estimation with information-guided selection and rejection of candidate interactions to recover both the structure and coefficients of Lorenz-63 \cite{trautner2020informative}. Unlike state assimilation alone, IEKL could revise the dynamical representation itself; unlike fitting weights in a fixed neural architecture, it optimized which candidate interactions were retained. Its recovered structure was informative within a specified library and selection criterion, not guaranteed globally optimal across arbitrary equations. Trautner, Li and Ravela subsequently showed how mathematical properties of polynomial dynamics could inform neural-network construction before training, supplying exact sizes for specified multiplicative networks and theory-guided bounds for other architectures \cite{trautner2021learn}. Physics could thus prescribe a network outright, guide its initial construction, or constrain subsequent informative adaptation.

Learning from incomplete physics also requires an account of what observations actually constrain. Ensemble Kalman approaches have trained hybrid physical--neural parameterizations using indirect observations without gradients \cite{lopezgomez2022ensemble}, while Ensemble-Conditional Gaussian Processes (Ens-CGP) provide a conditional-Gaussian interpretation of empirical ensembles and the covariance geometry linking physical parameters, observations and uncertainty under stated assumptions \cite{ravela2026enscgp}. These developments distinguish ensemble-conditioned inference from the numerical algorithms used to update parameters, and neither Gaussian conditioning nor an ensemble update alone selects a dynamical structure.

The need to combine physical knowledge with different kinds of inferred information is evident in geophysical applications. For rainfall downscaling, Saha and Ravela combined physics- and statistics-primed adversarial learning with ensemble-based optimal estimation for bias correction, using coarse-model information to reconstruct unresolved fine-scale rainfall and its distribution \cite{saha2024rainfall}. For cyclone intensification, Gee and Ravela instead identified an interpretable stochastic dynamical equation from observations and reanalysis, using sparsity promotion for equation-term selection and ensemble learning for coefficient refinement \cite{gee2026cyclone}. In magnetotelluric inversion, Kim, Ravela and Evans coupled Ens-CGP conditioning of physically admissible resistivity ensembles to a residual-learning neural network, propagating ensemble uncertainty through the corrected estimates \cite{kim2026npi}. Together these applications illustrate why uncertain physics may demand parameter estimation, learned fine-scale fields, inferred equations or constrained neural corrections, and why their treatment within one evolving operator graph is an architectural objective rather than a capability already established by any one application.

A remaining practical question is whether equation-derived neural execution can support substantial simulations rather than only illustrative dynamical systems. Todorovic, Gianchandani and Ravela addressed this in 2026 using the Korteweg--de Vries equation \cite{todorovic2026}. Their PolyNet reproduced the accuracy of the prescribed finite-difference Runge--Kutta method without training and was tested against conventional CPU and GPU solvers and three trained neural approaches, over integrations reaching $10^6$ resolution-dependent steps and ensembles reaching $10^4$ trajectories. Its GPU throughput was competitive, and its performance was comparable to an equivalent conventional GPU numerical graph. The result was that known physics can organize high-throughput tensor execution without first being learned. The unresolved test is whether this construction remains effective when the dynamics involve spherical geometry, unstructured staggered grids, rational operators, boundaries and forcing rather than a one-dimensional periodic polynomial equation.

The convergence of equation-derived neural execution, ensemble-based fluid inference, and data-informed geophysical modeling motivates Physics-Spawned Informative Networks (\phispin), conceived as a unified neurodynamical architecture in which known equations prescribe operator interactions, uncertain physical and neural parameters can be estimated together by adjoint-free ensemble inference, and eligible structural relationships can be selected and revised informatively. The complete coupled graph evolves through discrete-time neural integration; differentiability of its constituent operators is permitted but not required. The implementation of such graphs natively onto hardware is not, specifically, the subject of the architecture, but we note and use JAX for efficient GPU realizations.

The architecture is described next, followed by its first geophysical realization, \phispinsw. Motivated by eventual coastal flood modeling, we begin with the prescribed rotating shallow-water equations and test whether their full spatial and temporal computation can be executed directly through JAX on spherical Voronoi meshes. This study establishes the prescribed-computation component; the joint parameter and structural adaptation envisaged for the full architecture remains to be implemented for shallow-water dynamics.

\subsection{The \phispin\ architecture: prescribed and adaptive neurodynamics}
The central object in \phispin\ is a single neurodynamical operator graph whose degree of prescription or adaptation depends on the available knowledge. Some interactions and coefficients may be fixed exactly by governing equations; others may have known functional forms but uncertain physical parameters; still others may require data-derived functions or changes in operator structure. These are distinctions in what can be prescribed or inferred, not boundaries between a physical solver and an attached neural correction. In particular, parameter estimation and structural optimization can act on any eligible part of the coupled dynamics, including uncertain relationships conventionally described as physical. Only relationships treated as established constraints need remain fixed.

We express the complete discrete-time evolution as
\begin{equation}
\mathbf{x}^{n+1}
=\mathcal{T}^{\mathrm{neural}}_{\Delta t}
\left(\mathbf{x}^{n},\mathbf{u}^{n},\boldsymbol{\xi}^{n};
\boldsymbol{\alpha},\boldsymbol{\theta},\mathcal{S}\right),
\label{eq:phispin-architecture}
\end{equation}
where $\mathbf{x}^{n}$ is the state, $\mathbf{u}^{n}$ denotes external forcing or controls, and $\boldsymbol{\xi}^{n}$ represents stochastic inputs where present. The parameters $\boldsymbol{\alpha}$ and $\boldsymbol{\theta}$ denote physical and neural or other data-derived coefficients, respectively, while $\mathcal{S}$ specifies the operator structure. Any of these may contain prescribed and inferable elements. Crucially, $\mathcal{T}^{\mathrm{neural}}_{\Delta t}$ denotes the \emph{complete} neural evolution, including the coupled dynamical operators and the recurrent operations that advance them in time. For dynamics expressed continuously, the correspondence is
\begin{equation}
\frac{d\mathbf{x}}{dt}
=\mathcal{F}\left(\mathbf{x},t;
\boldsymbol{\alpha},\boldsymbol{\theta},\mathcal{S}\right),
\qquad
\mathcal{T}^{\mathrm{neural}}_{\Delta t}
=\mathcal{R}^{\mathrm{neural}}_{\Delta t}[\mathcal{F}],
\label{eq:phispin-continuous}
\end{equation}
where $\mathcal{R}^{\mathrm{neural}}_{\Delta t}$ is a recurrent neural integration circuit for a specified time-stepping method \cite{trautner2019}. This representation does not demand that every component originate as a smooth continuous-time tendency: discrete-time operators, stochastic dynamics, projections and state-dependent switches may also be included, with suitable coupling and integration rules. Nor is it restricted to an additive decomposition into ``physics'' and ``closure.'' The operators may share states, enter the same fluxes or constitutive relations, and interact multiplicatively or through constrained structural choices.

At one limit, every interaction and relevant coefficient is prescribed. The equations then determine the graph and no training is needed. For polynomial tendencies, the PolyNet construction supplies an exact representation of the specified operator through products and weighted sums; more general physical models can retain division, nonlinear constitutive laws, geometric transformations, masks and other non-polynomial operations directly. Neural integration executes this graph with the errors of its chosen numerical discretization and temporal method, not an additional fitted tendency error. This is the limit realized in the present shallow-water implementation.

At the other limit, inference is performed over the eligible uncertainties of the \emph{same} graph. Uncertain physical coefficients and neural parameters may be assembled into a joint inference vector,
\begin{equation}
\mathbf{p}
=\left[\boldsymbol{\alpha}_{\mathrm{uncertain}}^{\mathsf{T}},
\boldsymbol{\theta}_{\mathrm{uncertain}}^{\mathsf{T}}\right]^{\mathsf{T}},
\label{eq:phispin-parameters}
\end{equation}
and ensembles of forward neural integrations relate variations in $\mathbf{p}$ to variations in observed or diagnostically relevant outputs. This forward-ensemble perspective has precedents in efficient ensemble smoothing and experimental fluid assimilation \cite{ravela2007smoothing,ravela2010observatory}, but the intended inference here extends beyond state estimation to physical and neural parameters. Ens-CGP supplies the conditional-Gaussian representation of the empirical ensemble and the covariance geometry supporting this inference \cite{ravela2026enscgp}; ensemble Kalman learning provides the operational parameter updates \cite{trautner2020informative}. The same forward-model evaluations can train neural weights and estimate uncertain physical parameters without differentiating through the complete evolution. This does not make nonlinear inference exact or remove the limited-rank and identifiability constraints of finite ensembles, but it makes differentiability irrelevant as a prerequisite for learning: smooth operators remain admissible, and nondifferentiable operators need not be excluded.

The inference problem also extends beyond parameter values. IEKL alternates estimation with information-guided selection and rejection of candidate terms, permitting $\mathcal{S}$ to change as the evidence for particular interactions changes \cite{trautner2020informative}. Such adaptation can occur wherever the coupled graph admits structural uncertainty, rather than only in a data-driven closure. Its objective is informative structure selection within the supplied candidates, observations and physical constraints, not an assertion of unconstrained global optimality. The candidate representation can itself be informed before learning by physics-derived PolyNet constructions and learnability-based sizing \cite{trautner2021learn}. Inferred interactions may remain flexible neural operators or be rendered as interpretable equations, as illustrated by the cyclone equation-discovery study \cite{gee2026cyclone}. In this way, prescribed, estimated and structurally selected relationships retain their distinct epistemic roles while contributing to one dynamical evolution.

The relation between operators and executable graphs makes the prescribed and adaptive limits computationally compatible.  When the governing equations are prescribed, they can be implemented directly as JAX computational graphs and compiled for GPU execution. When model adaptation is required, the computational graph can be lifted to an operator-aware representation (such as PolyNets or other types of Tensor graphs) in which physical and learned operators, their parameters, and their interactions can be modified through informative learning. The augmented model can then be translated back into an executable graph. This correspondence requires preserving operator identities and discretization provenance, rather than attempting to recover governing equations from optimized executable code. Structural changes may require recompilation. Thus, \phispin~accommodates both direct execution of prescribed physics and adaptive modification of the underlying operators, independently of the computational framework used. JAX is the execution technology used here, not a restriction on the architecture.

The distinction from existing hybrid models lies in the scope of inference. NeuralGCM \cite{kochkov2024neural} learns differentiable parameterizations within prescribed atmospheric dynamics but does not adapt their operator structure. UDEs \cite{rackauckas2020ude} couple mechanistic and learned dynamics and permit symbolic equation recovery, but do not provide joint, information-guided optimization of operator structure and parameters. Trautner, Margolis, and Ravela \cite{trautner2020informative} introduced precisely this capability through adjoint-free Informative Ensemble Kalman Learning (IEKL), while Gee and Ravela \cite{gee2026cyclone} demonstrated a complementary approach to identifying governing equations directly from observations. \phispin\ unifies prescribed full-graph neural integration and accelerator execution with informative adaptation of operator structure and joint physical and neural parameter inference, without requiring differentiability. The present work demonstrates the prescribed execution pathway at geophysical scale; the complete adaptive learning loop remains to be implemented and evaluated.

\subsection{Prescribed shallow-water realization of \phispin-SW}
Rotating shallow-water dynamics provide a stringent next step beyond the polynomial Korteweg--de Vries example and a necessary foundation for the coastal flood applications that motivate this work. The model must combine nonlinear transport and rotation with a spherical geometry, unstructured cell connectivity, staggered prognostic variables, non-polynomial potential-vorticity operations and, in coastal configurations, bathymetry, boundary masks and external forcing. These features test whether the governing equations can continue to prescribe the executed graph when its operators are no longer a simple polynomial stencil. We begin from the vector-invariant rotating shallow-water equations,
\begin{align}
\frac{\partial h}{\partial t}
&= -\nabla\cdot(h\mathbf{u}), \label{eq:continuity}\\
\frac{\partial\mathbf{u}}{\partial t}
&= -(\zeta+f)\,\hat{\mathbf{r}}\times\mathbf{u}
-\nabla(K+g\eta)+\mathbf{F}.
\label{eq:momentum}
\end{align}
Here, $h$ is layer thickness and $\mathbf{u}$ is velocity tangent to the sphere. Continuity expresses the negative divergence of transport $h\mathbf{u}$, while the momentum equation combines the rotational acceleration from absolute vorticity $\zeta+f$, the gradient of kinetic and gravitational potential energy, and forcing or friction $\mathbf{F}$. The relative vorticity is $\zeta=\hat{\mathbf{r}}\cdot(\nabla\times\mathbf{u})$, with $\hat{\mathbf{r}}$ the outward radial unit vector; $f=2\Omega\sin\phi$ is the Coriolis parameter at latitude $\phi$ for rotation rate $\Omega$. The kinetic energy per unit mass is $K=|\mathbf{u}|^2/2$, and the free-surface elevation is $\eta=h+b$, where $b$ is fixed bottom elevation measured positive upward. Spatial derivatives act tangentially to the sphere.

The physical terms already indicate how the operator graph is assembled. On a fixed grid, spatial differentiation is a weighted combination of neighboring degrees of freedom. For example,
\begin{equation}
\left.\frac{\partial u}{\partial x}\right|_i
\approx \sum_j w_{ij}^{(1)}u_j,
\end{equation}
with weights $w_{ij}^{(1)}$ prescribed by geometry and the derivative formula. Nonlinear advection is then represented by multiplication of known fields and a prescribed spatial operation,
\begin{equation}
u_i\left.\frac{\partial u}{\partial x}\right|_i
\approx \sum_j w_{ij}^{(1)}u_i u_j.
\end{equation}
This quadratic interaction is a PolyNet operation. The derivative's spatial order is independent of the polynomial degree: constant-diffusivity diffusion, for instance, has the linear form
\begin{equation}
\left.\nu\frac{\partial^2u}{\partial x^2}\right|_i
\approx \nu\sum_j w_{ij}^{(2)}u_j.
\end{equation}
Diffusion serves only to illustrate this distinction and is not part of the inviscid equations above.

\begin{figure}[!hb]
\centering
\includegraphics[width=\linewidth]{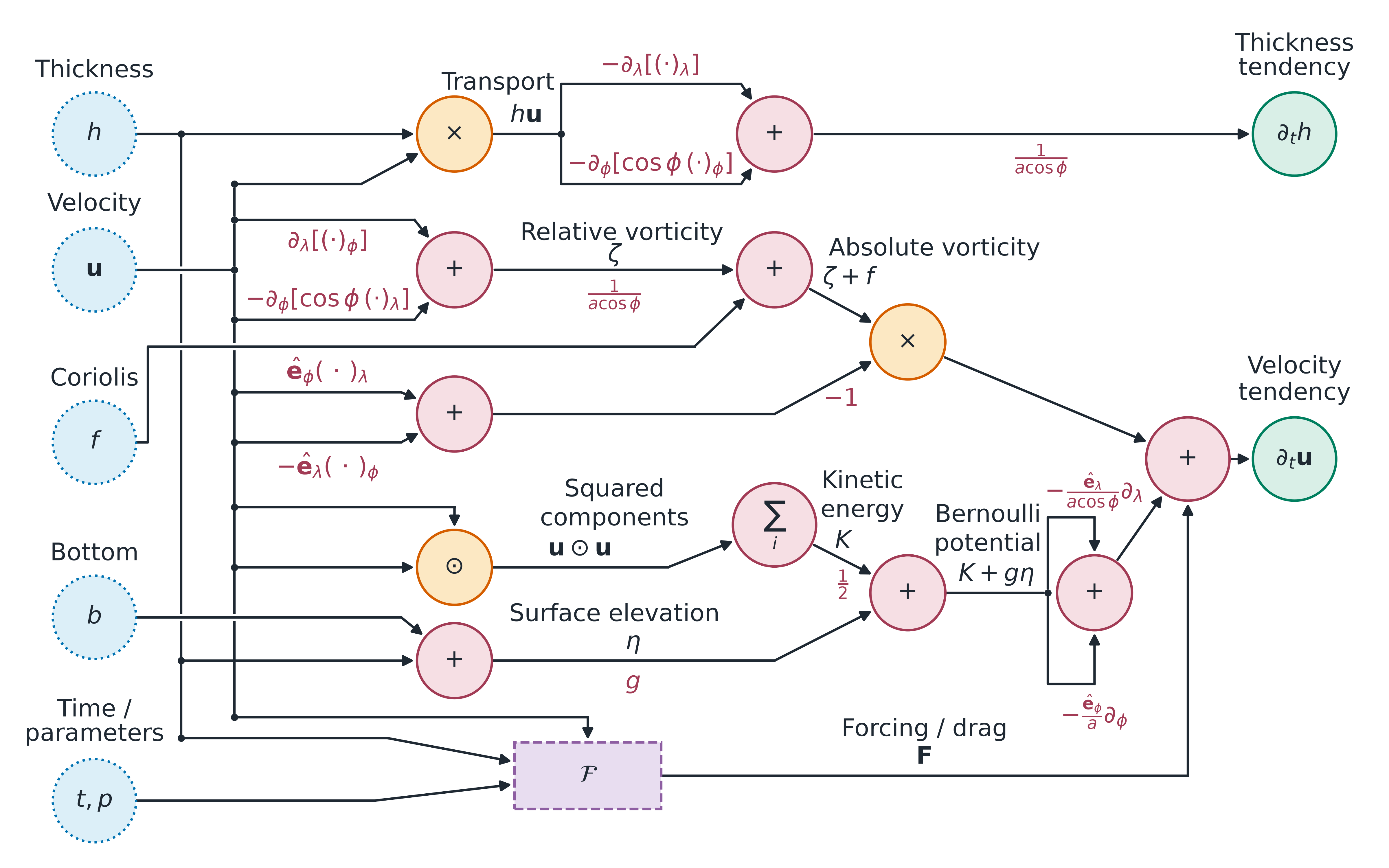}
\caption{\textbf{Computational graph of the rotating shallow-water equations.} The equation-derived graph evaluates the instantaneous tendencies $\partial_t h=-\nabla\cdot(h\mathbf{u})$ and $\partial_t\mathbf{u}=-(\zeta+f)(\hat{\mathbf{r}}\times\mathbf{u}) -\nabla(K+g\eta)+\mathbf{F}(h,\mathbf{u},t;p)$, where $\eta=h+b$, $K=|\mathbf{u}|^2/2$, and $\zeta=\hat{\mathbf{r}}\cdot(\nabla\times\mathbf{u})$. Here, $h$ is layer thickness, $\mathbf{u}$ horizontal velocity, $b$ positive-up bottom elevation, $g$ gravitational acceleration, $f$ the Coriolis parameter, and $\hat{\mathbf{r}}$ the outward radial unit vector. Derivatives act tangentially to the sphere. Dotted blue circles denote inputs; green circles denote tendencies; rose circles denote additions or component summation; and amber circles denote products. Componentwise multiplication ($\odot$) and summation over the two orthonormal tangent-velocity components, indexed by $i$, construct $K$. Magenta labels show fixed weights, including spatial derivative operators, geometric factors, basis transformations, and signs. Here, $\lambda$ and $\phi$ denote longitude and latitude, $a$ is the sphere radius, and $\hat{\mathbf{e}}_\lambda$ and $\hat{\mathbf{e}}_\phi$ are the local eastward and northward unit vectors. Component subscripts select the corresponding velocity components. Arrows carry intermediate fields; black dots mark shared branches, while unmarked crossings are unconnected. The separate dashed purple rectangle represents unexpanded, optional, possibly non-polynomial forcing or drag, evaluated from the state, time $t$, and prescribed parameters $p$. The five internal columns organize elementary operations by their earliest dependency level, rather than physical or learned layers. This continuum decomposition does not expand discrete C-grid operators or numerical stabilization.}
\label{fig:swe-computational-graph}
\end{figure}

In shallow water, the flux $h\mathbf{u}$, kinetic energy $K$ and rotational interaction $(\zeta+f)\hat{\mathbf{r}}\times\mathbf{u}$ similarly prescribe products, sums and spatial operators. The continuity branch constructs and takes the negative divergence of mass transport; the momentum branch combines rotation, the gradient of $K+g\eta$, and $\mathbf{F}$. Figure~\ref{fig:swe-computational-graph} displays the resulting continuum-level dependencies as a physically interpretable operator graph. Its nodes indicate operations and shared intermediate fields, rather than learned layers; the C-grid discretization supplies the detailed sparse connectivity and geometric coefficients used in execution.

The actual execution graph is determined by the compatible spherical C-grid discretization, which places layer thickness at Voronoi cell centers and normal velocity at edges while diagnosing vorticity and potential vorticity at dual vertices. In particular, potential vorticity involves division of absolute vorticity by interpolated thickness. \phispinsw\ retains that division exactly, along with configured non-polynomial forcing, friction, masks and protective operations, instead of replacing them with polynomial or trained approximations (Methods). 
\begin{figure}[!hb]
\centering
\includegraphics[width=\linewidth]{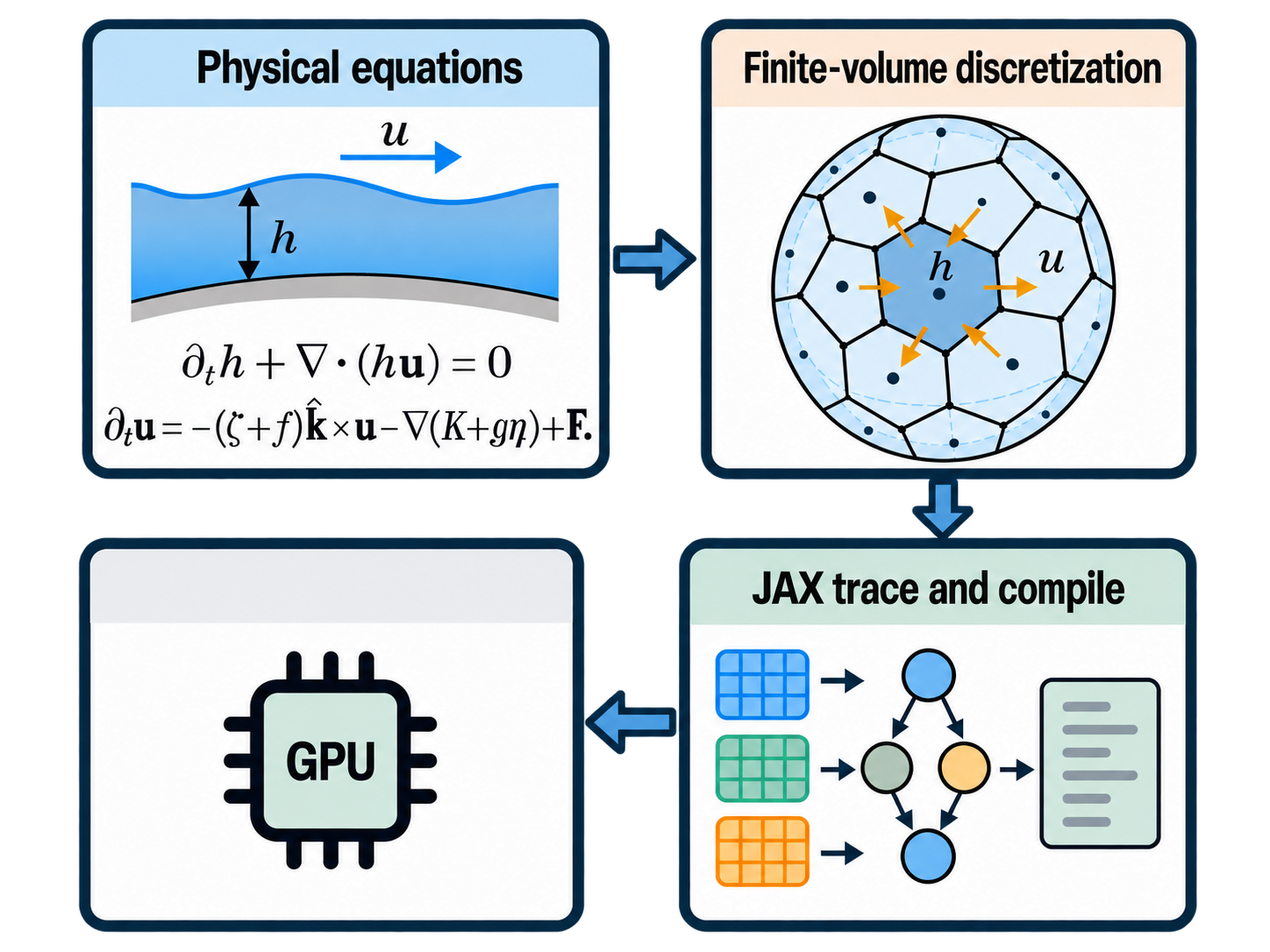}
\caption{\textbf{From governing equations to GPU execution.}
Clockwise from the upper left: the rotating shallow-water equations specify the evolution of layer thickness and horizontal velocity. C-grid discretization on an unstructured spherical grid places thickness at cell centres and normal velocity at edges, producing a finite system of evolution equations. JAX traces the prescribed array operations and compiles them through XLA for execution on the GPU. The compiled model advances the state using the specified recurrent neural-integration scheme, with parallel calculations across the grid. The governing physics, operator structure and integration coefficients are prescribed, not learned. This figure shows the implemented direct-execution pathway, not the proposed adaptive reconstruction and learning pathway.}
\label{fig:workflow}
\end{figure}
The construction thus extends the prescribed PolyNet principle to the richer operator algebra required by shallow-water dynamics. Figure~\ref{fig:workflow} follows this discretized graph into accelerator execution.

Time advancement is part of the same network. In the neural-integration construction, each Runge--Kutta stage reuses the prescribed tendency graph and combines intermediate states with fixed weights \cite{trautner2019,todorovic2026}. For classical fourth-order Runge--Kutta (RK4), the complete discrete-time update is
\begin{equation}
\mathbf{y}^{n+1}
=\mathbf{y}^{n}
+\frac{\Delta t}{6}
\left(\mathbf{k}_1+2\mathbf{k}_2
+2\mathbf{k}_3+\mathbf{k}_4\right),
\end{equation}
where $\mathbf{y}$ contains the discrete prognostic fields. The four tendencies are evaluated at provisional states at physical times $t$, $t+\Delta t/2$, $t+\Delta t/2$ and $t+\Delta t$; each provisional state is a prescribed combination of the current state and an earlier stage tendency. Hence the operator and its recurrence jointly define the neural time evolution. \phispinsw\ implements these equivalent RK4 computations directly as JAX operations, preserving the selected method's spatial and temporal errors without introducing a fitted-tendency approximation. The description as neural integration does not assert a different RK4 numerical method.

For execution, geometric coefficients and mesh connectivity are prepared once, neighboring fields are gathered through sparse indexing, and sums, products and other physical operations are evaluated as JAX arrays \cite{jax2018github}. XLA compiles the traced calculations for individual GPUs, fusing compatible operations and avoiding dense global spatial matrices. Parallel grid operations can therefore exploit the accelerator, while the dependencies between RK4 stages and successive timesteps remain sequential. This is the direct equations-to-executable-graph path: no separately materialized PolyNet and no training step intervene. We specified the model formulation, numerical methods and verification criteria and used OpenAI Codex to assist with implementation, debugging, testing and documentation (Methods). The experiments below evaluate this prescribed realization against analytical and MPAS references and measure its GPU performance; ensemble-conditioned training and informative structural adaptation remain the next architectural step.

\section{Results}
\subsection{Numerical Accuracy and intermodel agreement}
We assessed \phispinsw\ using Williamson tests 1-5, covering prescribed transport, balanced flows, time-dependent forcing and nonlinear flow over topography. Tests 1--4 provide analytical references, while test 5 compares the evolving solution with the shallow-water model of the Model for Prediction Across Scales (MPAS) framework, denoted as MPAS-SW. The G6 simulations reproduce the balanced jet in test 3 and the translating low in test 4, while cosine-bell transport in test 1 exhibits larger relative errors (Supplementary Text).

\begin{figure}[!ht]
\centering
\includegraphics[width=\linewidth]{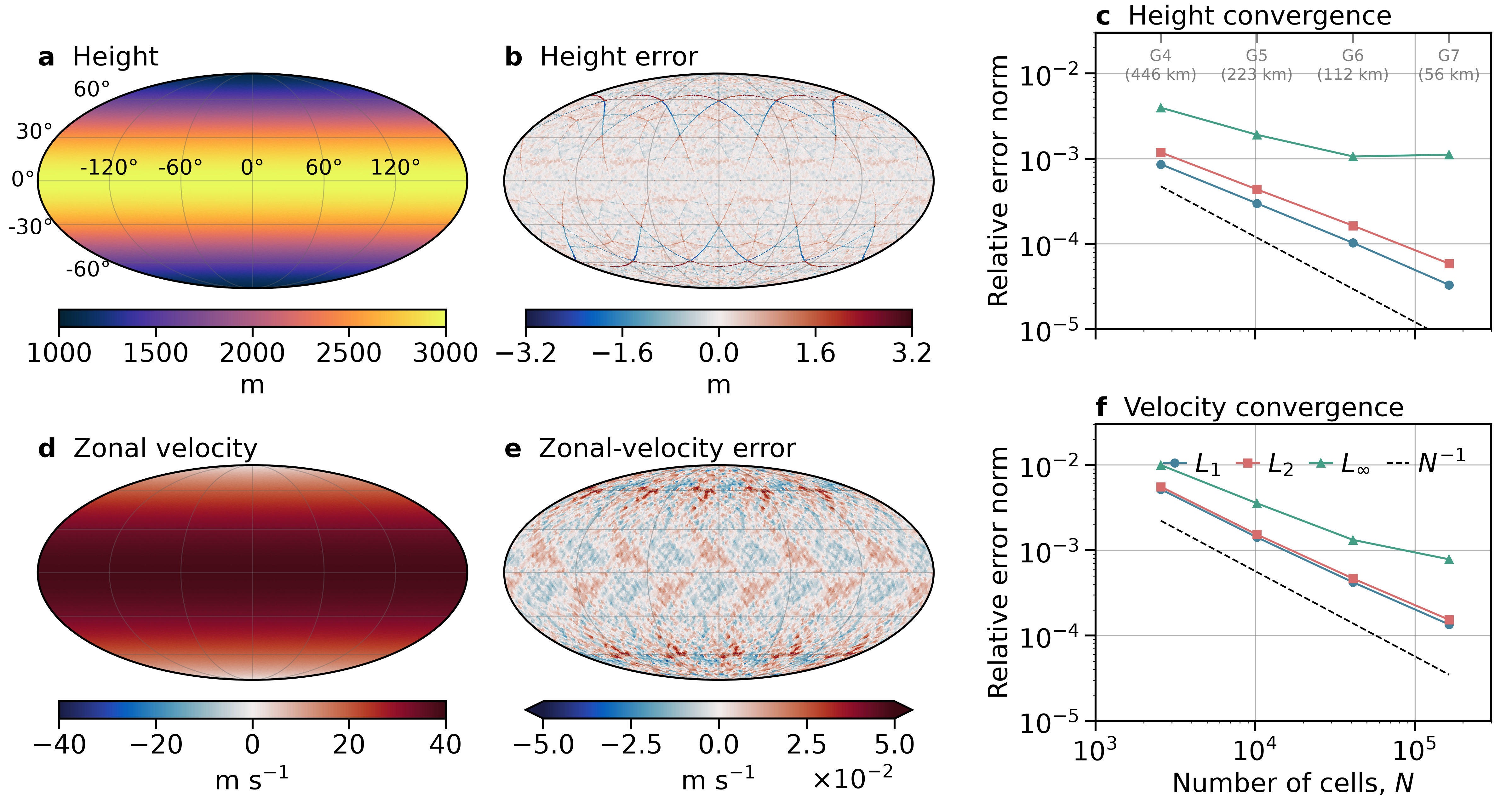}
\caption{
\textbf{Williamson test 2: steady geostrophic flow.}
Native-grid G6 height (a) and reconstructed zonal velocity (d)
near day 5, with numerical-minus-analytical errors in (b,e).
Panels (c,f) show relative $L_1$, $L_2$ and $L_{\infty}$ error
norms for height and full-vector velocity across G4--G7,
plotted against cell count, $N$. The black $N^{-1}$ reference
lines indicate second-order scaling with grid spacing on
these quasi-uniform grids.
}
\label{fig:williamson2}
\end{figure}

Williamson test 2 reveals how numerical error changes with grid refinement. Its prescribed flow is steady and geostrophically balanced, so departures from the analytical solution measure error accumulated during integration. We compared solutions near day 5 across grids G4 (2562 cells, $\Delta l\sim446~km$) to G7 (163842 cells, $\Delta l\sim56~km$). Panels (a,d) in Fig.~\ref{fig:williamson2} show height and reconstructed zonal velocity alongside their numerical-minus-analytical errors in panels (b,e) on G6 (40962 cells, $\Delta l\sim112~km$). Relative $L_1$, $L_2$ and $L_\infty$ norms quantify these errors (Fig.~\ref{fig:williamson2}c,f): the cell-area-weighted mean absolute error, cell-area-weighted root-mean-square error and maximum pointwise error, respectively. Each uses the corresponding analytical norm for normalization. Velocity norms evaluate the full horizontal vector, although the maps show only its zonal component.

Across G4–G7, the relative $L_2$ height error decreases from $1.2\times10^{-3}$ to $5.9\times10^{-5}$, while the relative $L_2$ velocity error decreases from $5.6\times10^{-3}$ to $1.5\times10^{-4}$. These approximately 20-fold and 36-fold reductions show substantial improvement in domain-wide accuracy. However, the relative height $L_\infty$ error remains nearly constant from G6 to G7, indicating that refinement does not consistently reduce the largest local discrepancy. Similar maximum-norm stagnation has been reported for TRiSK-based shallow-water discretizations \cite{yu2020comparing,calandrini2021comparing}. Modified formulations can improve $L_\infty$ convergence \cite{peixoto2016accuracy}, but evaluating alternative spatial schemes is beyond the scope of this work. Because the time step also decreases with mesh spacing, this comparison measures coupled spatial and temporal refinement. The reported endpoint reductions correspond to effective spatial orders of approximately 1.45 for height and 1.74 for velocity, rather than a demonstration of exact second-order convergence; the $N^{-1}$ lines are reference slopes.

Williamson test 5 considers zonal flow over an isolated mountain. We use this test to assess whether \phispinsw's fixed-weight computational graph implementation on the GPU preserves the evolution produced by MPAS-SW as the flow becomes increasingly nonlinear. Surface height characterizes the large-scale wave response, while PV provides a complementary diagnostic of the deformation and transport of vortical structures.

\begin{figure}[!ht]
\centering
\includegraphics[width=\linewidth]{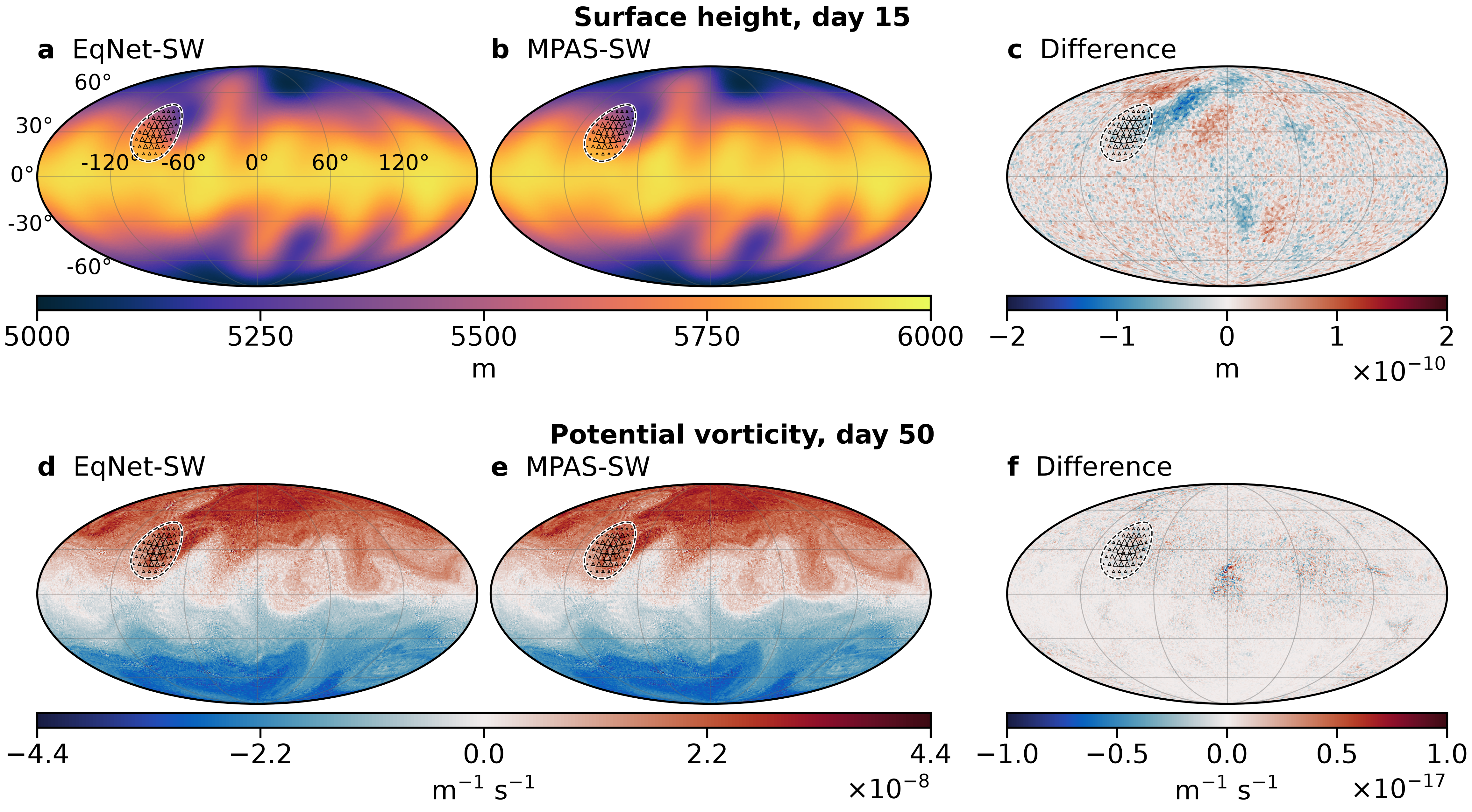}
\caption{
\textbf{Williamson test 5: flow over an isolated mountain.} Native-grid surface height $\eta$ at day 15 (a–c) and potential vorticity at day 50 (d–f). Columns show \phispinsw, MPAS-SW, and their difference (\phispinsw\ minus MPAS-SW). Both simulations use the same G6 grid (40,962 cells, typical resolution of $\sim 112~km$) and a $150~s$ timestep. Dashed outlines and hollow triangles indicate the prescribed mountain. Differences quantify intermodel agreement.}
\label{fig:williamson5}
\end{figure}

Both models use the same native G6 grid (40,962 cells, typical resolution of $\sim112\,\mathrm{km}$), initial conditions, topography, and 150-$s$ time step, and are integrated for 50 days. \phispinsw\ and MPAS-SW produce visually indistinguishable surface-height patterns at day 15 (Fig.~\ref{fig:williamson5}a,b) and PV distributions at day 50 (Fig.~\ref{fig:williamson5}d,e). The maximum pointwise differences are $1.85\times10^{-10}\,\mathrm{m}$ for surface height (Fig.~\ref{fig:williamson5}c) and $9.74\times10^{-18}\,\mathrm{m^{-1}\,s^{-1}}$ for PV (Fig.~\ref{fig:williamson5}f). Both fields are compared on their native supports without spatial interpolation and the differences quantify disagreement between the implementations, rather than independent accuracy against observed physical conditions.

The saved conservation histories also closely coincide (Supplementary Fig. \ref{fig:w5-conservation}). Relative energy loss reaches approximately $10^{-9}$. Potential enstrophy increases by approximately 2\% in both simulations over 50 days. Together, these results support the fidelity of the computational-graph implementation to the reference solver through extended nonlinear evolution, while the shared enstrophy drift remains a property of the selected fully discrete numerical formulation.

\subsection{Hardware performance and energy efficiency}
We assess \phispinsw's computational performance using Williamson test~5. At each refinement from G5 to G9, \phispinsw\ and MPAS-SW advance the same initial state for 2000 steps using matched grids and timesteps. Each trial restarts from that initial state. We measure integration time and read cumulative hardware energy counters around the integration phase: CPU-package counters for MPAS-SW and NVIDIA GPU-board counters for \phispinsw. The difference between the initial and final counter readings gives the energy consumed. Throughput is the number of completed steps divided by integration time, while productivity is the number of completed steps divided by energy consumed. Compilation and initialization are excluded. CPU energy measurements include timestep-loop bookkeeping beyond the narrower native integration timer. Markers show medians and bars show min--max ranges across ten independent trials, including the first execution after compilation. The timestep decreases with refinement, so comparisons use matched workloads at each resolution rather than a common simulated duration across resolutions.

\begin{figure}[!ht]
\centering
\includegraphics[width=\linewidth]{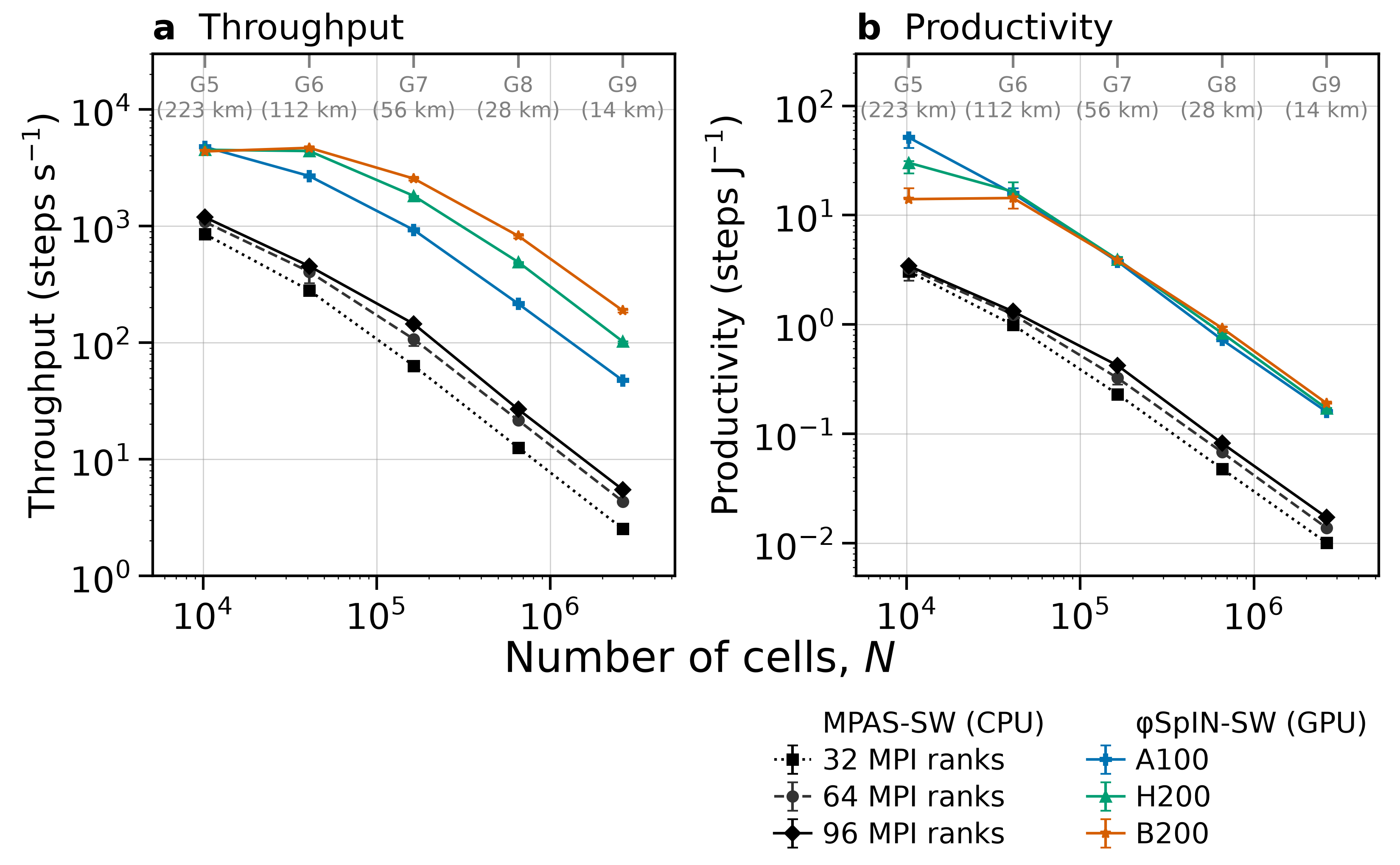}
\caption{\textbf{Computational performance of \phispinsw\ and MPAS-SW.} (a) Integration throughput (steps/s) and (b) productivity (steps/J) for Williamson test~5 across refinements G5--G9. \phispinsw\ runs on individual NVIDIA A100, H200 and B200 GPUs; MPAS-SW uses 32, 64 or 96 MPI ranks on one socket of an AMD EPYC 9654 processor. Markers show medians and bars show min--max ranges across ten independent 2000-step trials. All runs use float64 precision, APVM disabled, matched grids, initial states and refinement-dependent timesteps. Compilation is excluded. Energy covers the CPU package or GPU board, excluding the GPU host and facility cooling.}
\label{fig:speed-benchmarks}
\end{figure}

At G9, containing approximately 2.6 million cells, \phispinsw\ achieves 9-fold, 19-fold and 34-fold higher throughput on the A100, H200 and B200, respectively, compared with MPAS-SW using 96 cores of an AMD EPYC 9654 processor (Fig.~\ref{fig:speed-benchmarks}a). The corresponding GPU productivities are approximately 0.16, 0.17 and 0.19~steps~J$^{-1}$, compared with 0.010, 0.014 and 0.017~steps~J$^{-1}$ for MPAS-SW using 32, 64 and 96 cores, respectively (Fig.~\ref{fig:speed-benchmarks}b). Relative to the 96-core configuration, these represent approximately $9\times$, $10\times$ and $11\times$ improvements. Across G6--G9, all three GPUs deliver approximately an order-of-magnitude higher productivity than the 96-core configuration.

Note that energy measurements exclude the GPU host and facility cooling and therefore do not represent complete-system consumption. Longer trials on the A100 at G6 yielded essentially unchanged median throughput and productivity. These benchmarks establish an advantage over the specified CPU implementation and hardware measurement boundaries, not an intrinsic speed or energy advantage over equivalently optimized GPU solvers or a complete-system energy comparison.

\subsection{Idealized cyclone response on a variable-resolution grid}
\begin{figure}[!ht]
\centering
\includegraphics[width=0.95\linewidth]{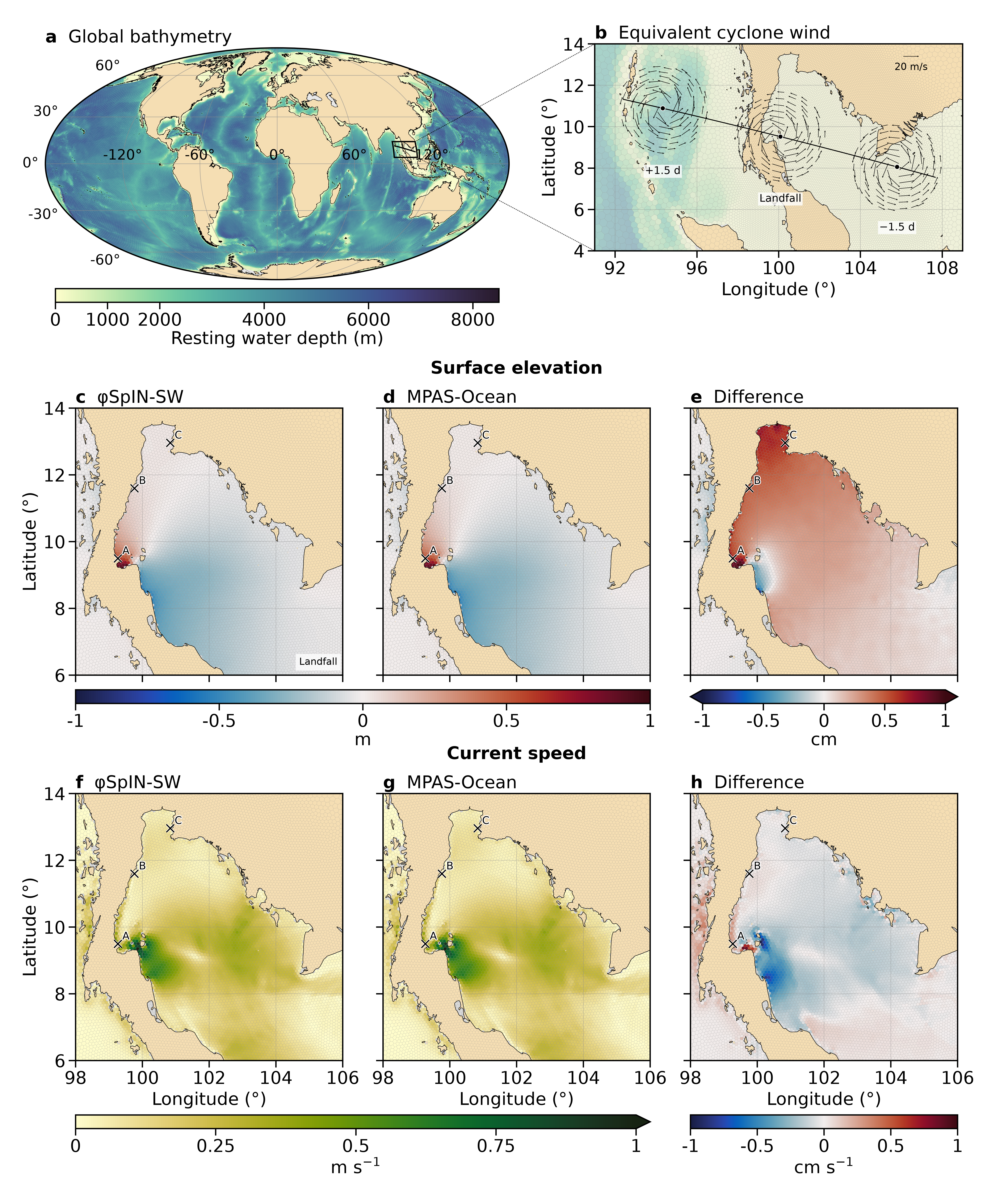}
\caption{\textbf{Idealized cyclone response in the Gulf of Thailand.} (a) Global bathymetry  with the study area boxed. (b) Variable grid, cyclone track and equivalent winds at $-1.5$, $0$ and $+1.5$ days from landfall. (c,d) Surface elevation and (f,g) reconstructed current speed at landfall from \phispinsw\ and MPAS-Ocean; (e,h) their signed differences (\phispinsw\ minus MPAS-Ocean). Crosses A--C mark coastal samples. Both models share physical inputs and prescribed stress samples; equivalent winds in (b) are for visualization only.}
\label{fig:thai-cyclone}
\end{figure}

To complement the performance benchmarks, we examine how \phispinsw\ accommodates a coastal application: an idealized cyclone crossing the Gulf of Thailand. The experiment combines variable bathymetry, an irregular coastline and local grid refinement (Fig.~\ref{fig:thai-cyclone}a,b). These coastal conditions motivate using single-layer MPAS-Ocean rather than MPAS-SW as the comparison model. Its existing coastal boundary, bottom-friction and drying-prevention capabilities accommodate this experiment without source-code modifications; using MPAS-SW would require additional coastal treatments. Single-layer MPAS-Ocean configurations have previously been used to simulate Hurricane Sandy storm surge \cite{lilly2023,lilly2025}, compound river–coastal flooding during Hurricane Irene \cite{feng2024}, and coastal inundation with subgrid corrections \cite{brus2026}.

\phispinsw\ and single-layer MPAS-Ocean share the same bathymetry, fixed coastline and initially resting state. Both receive the same cyclone-induced wind-stress samples, with no background wind stress, and use explicit quadratic bottom drag with a constant coefficient $C_D=0.0025$. The simulations span two days before to two days after landfall, defined as $t=0$, with a $2$~s time step. The prescribed cyclone translates westward across the Gulf; Fig.~\ref{fig:thai-cyclone}b shows its track and circulation at $t=-1.5$, $0$ and $+1.5$ days. The equivalent wind vectors illustrate the prescribed stress forcing and are not model inputs.

At landfall, both models produce coastal setup near the landfall region and a broader surface depression across the central Gulf (Fig.~\ref{fig:thai-cyclone}c,d). Their reconstructed current-speed fields show closely corresponding circulation patterns, including a strong coastal response and currents approaching $1~m~s^{-1}$ (Fig.~\ref{fig:thai-cyclone}f,g). The difference maps reveal discrepancies that are difficult to distinguish in the model fields themselves: surface-elevation differences extend across the basin, while the larger current-speed differences are localized near the coast (Fig.~\ref{fig:thai-cyclone}e,h).

\begin{figure}[!ht]
\centering
\includegraphics[width=\linewidth]{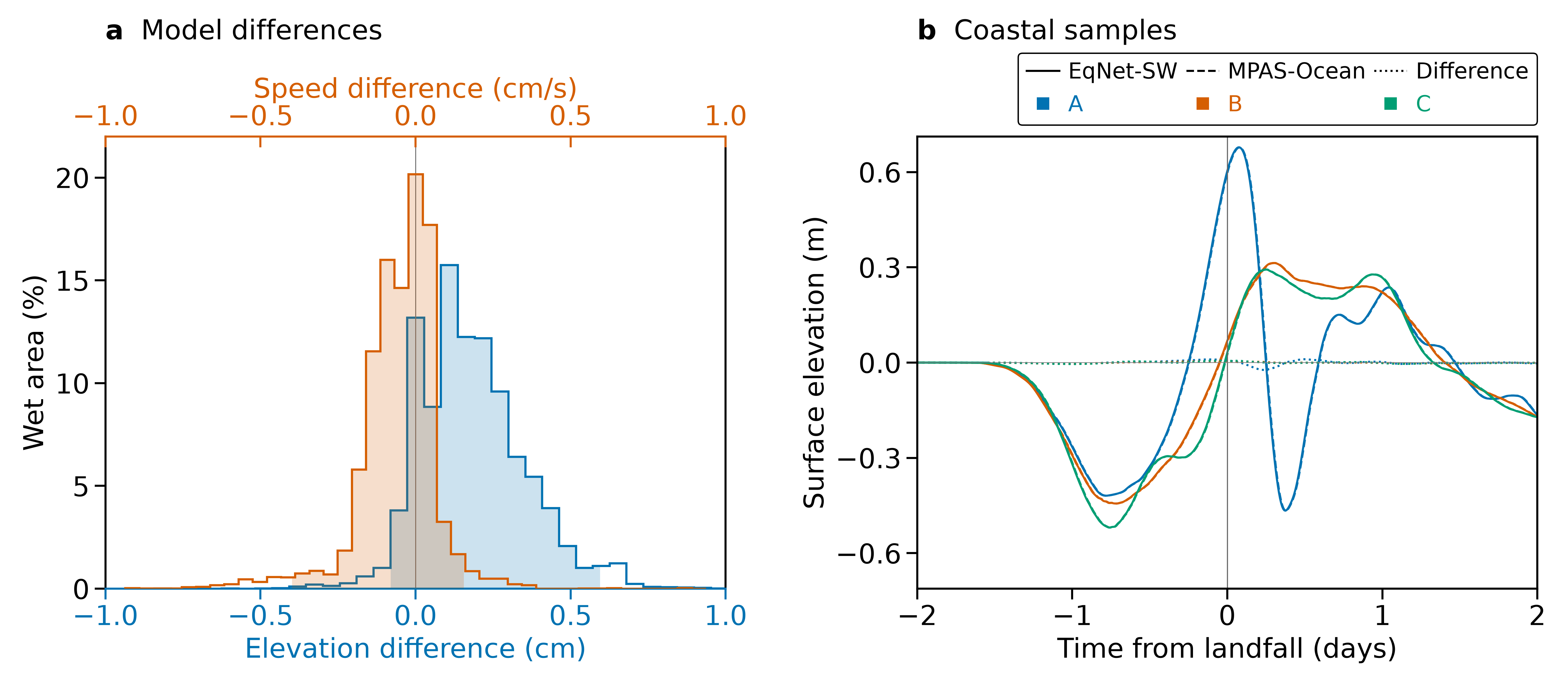}
\caption{\textbf{Cross-model differences and coastal cyclone response.} (a) Distributions of signed \phispinsw\ minus MPAS-Ocean differences at landfall: surface elevation in cm (blue, bottom axis) and current speed in cm~s$^{-1}$ (orange, top axis). Each bin represents a percentage of the native wet-cell area within the mapped Gulf. Shading identifies the central area-weighted 95\% intervals. (b) Surface-elevation histories at coastal locations A--C marked in Fig.~\ref{fig:thai-cyclone}. Colors identify locations; solid and dashed lines show \phispinsw\ and MPAS-Ocean, respectively, and dotted lines show their signed differences. Landfall occurs at $t=0$.}
\label{fig:thai-cyclone-diagnostics}
\end{figure}

The distributions in Fig.~\ref{fig:thai-cyclone-diagnostics}a quantify these differences over the same region. Histogram bins report percentages of native wet-cell area, accounting for the variable grid resolution. At landfall, the central 95\% intervals of signed \phispinsw\ minus MPAS-Ocean differences are $[-0.0804,0.595]~cm$ for surface elevation and $[-0.398,0.156]~cm~s^{-1}$ for current speed. These shaded intervals are bounded by the area-weighted 2.5th and 97.5th percentiles. The differences within these intervals are small relative to the metre-scale elevation response and $\sim1~m~s^{-1}$ currents, but do not imply a uniform percentage difference at every location.

The coastal histories show that agreement persists through the four-day evolution (Fig.~\ref{fig:thai-cyclone-diagnostics}b). At locations A--C, both models capture the pre-landfall drawdown, subsequent rise and post-landfall adjustment. Location A has a sharper response, reaching approximately $0.68~m$, whereas B and C develop broader positive elevations approaching $0.3~m$. The model trajectories largely overlap, although transient differences at A reach approximately $2.3~cm$.

Although the models share the same physical setup and forcing samples, differences in numerical implementation may account for the small discrepancies in their responses. For instance, \phispinsw\ linearly interpolates the prescribed stress between samples at RK4 stage times, whereas MPAS-Ocean holds each sample constant until the next forcing update. \phispinsw\ uses a fixed coastal mask and MPAS-Ocean a land-culled mesh; their drying-prevention treatments also remain model-specific.

Nevertheless, \phispinsw\ captures the principal cyclone-driven response without calibration to MPAS-Ocean trajectories or observational data. This agreement establishes consistency between the models under prescribed forcing; predictive accuracy against observations and coastal inundation require separate assessment. The following section discusses the broader implications of equation-derived computational graphs for representing and executing physics in Earth system models.

\section{Summary and Discussion}

In this work, we developed and demonstrated \phispinsw,
a training-free, equation-derived neural realization of the
rotating shallow-water equations on unstructured spherical
Voronoi meshes. We incorporated
compatible C-grid discretization, geometric operators,
nonlinear transport, rational potential-vorticity operations,
forcing, friction and coastal boundary treatments within
a unified computational network. Recurrent Runge--Kutta
integration advances the complete network in discrete time,
with JAX providing efficient GPU execution. This extends
the earlier demonstration of PolyNets through
the Korteweg--de Vries equation to complex global and
coastal geophysical flows.

Analytical Williamson benchmarks established the numerical
accuracy and refinement behavior of the prescribed scheme.
In the nonlinear mountain-flow experiment, \phispinsw\
reproduced MPAS-SW over 50 days, with maximum surface-height
and potential-vorticity differences of
$1.85\times10^{-10}~m$ and
$9.74\times10^{-18}~m^{-1}~s^{-1}$, respectively.
At approximately $14~km$ global resolution, containing
2.6 million cells, the A100, H200 and B200 implementations
achieved 8.6-, 19- and 34-fold higher timestep throughput
than the 96-core MPAS-SW CPU reference, together with
approximately 9--11-fold improvements in measured
device-level energy productivity. An idealized cyclone
experiment in the Gulf of Thailand further demonstrated
close agreement with MPAS-Ocean in coastal water elevations
and currents on a variable-resolution mesh. Together,
these experiments establish the fidelity, computational
scalability and applicability of prescribed neural
integration to substantially more complex geophysical
dynamics, providing the computational foundation for
subsequent adaptive \phispin\ implementations.

The construction of \phispinsw\ follows the PolyNet formulation
of neural integration \cite{trautner2019,todorovic2026}.
Figure~\ref{fig:swe-computational-graph} extends the
continuous-time PolyNet representation to the rotating
shallow-water equations, expressing their governing interactions
as a network of physically interpretable operators.
Spatial discretization preserves this operator representation,
with differential operators realized as prescribed,
geometry-dependent weighted connections and nonlinear
interactions represented through products and other elementary
operations. The PolyNet construction generalizes to an
Operator Network (OpNet), accommodating the rational and
non-polynomial operations required by shallow-water dynamics.
Recurrent Runge--Kutta time stepping then produces the complete
discrete-time OpNet, incorporating both the dynamical operators
and their neural integration. JAX realizes this discrete-time
network as an executable tensor graph, compiling the prescribed
operations for efficient GPU execution. This realization can
proceed directly from the governing equations to JAX without
requiring the intermediate OpNet to be explicitly materialized.

This correspondence also provides the computational pathway
for adaptive \phispin. With operator identities, geometric
relationships and discretization metadata preserved, the
JAX computational graph can be lifted back into its
continuous-operator OpNet representation. This representation
can then be augmented with data-derived neural components,
learned equations and structurally adaptable interactions.
Ens-CGP-based ensemble inference can estimate uncertain
physical parameters and neural weights, while IEKL can
informatively select and optimize eligible operator structure
throughout the coupled graph
\cite{ravela2026enscgp,trautner2020informative}.

The adapted OpNet is then returned to discrete-time neural
integration and compiled through JAX for GPU execution.
This process can be repeated through successive ensemble-learning
and structural-adaptation cycles. Both differentiable and
nondifferentiable operators are admissible, and learned components
may represent continuous-time dynamics or discrete-time operations.

The present work establishes and validates the prescribed
OpNet realization for rotating shallow-water dynamics.
Subsequent implementations will incorporate ensemble-based
parameter estimation and informative structural adaptation
within the same operator and execution framework. These
developments will require evaluation of parameter and structural
identifiability, ensemble-rank limitations, numerical stability,
conservation properties, and computational costs across
adaptation cycles. For coastal flood modeling, learned and
adapted interactions can be introduced at physically meaningful
cell, edge and vertex operators, with appropriate conservation
constraints and moving wet--dry boundary treatments.

We posit that the resulting architecture preserves established
physics executing
the system through neural integration, estimates uncertain parameters, and informatively
adapts incomplete dynamics without adjoints irrespective of the differentiability of its operators.
Comparisons with NeuralGCM, Universal Differential Equations,
and existing adjoint-free hybrid calibration methods will assess
the performance and predictive value of this approach
\cite{kochkov2024neural,rackauckas2020ude,lopezgomez2022ensemble}.

\clearpage 

\bibliography{ref}

@article{sridhar2022large,
  title="{Large-eddy simulations with ClimateMachine v0. 2.0: a new open-source code for atmospheric simulations on GPUs and CPUs}",
  author={Sridhar, Akshay and Tissaoui, Yassine and Marras, Simone and Shen, Zhaoyi and Kawczynski, Charles and Byrne, Simon and Pamnany, Kiran and Waruszewski, Maciej and Gibson, Thomas H and Kozdon, Jeremy E and others},
  journal={Geoscientific Model Development},
  volume={15},
  number={15},
  pages={6259--6284},
  year={2022},
  publisher={Copernicus Publications, G{\"o}ttingen, Germany}
}

@article{hafner2021fast,
  title="{Fast, cheap, and turbulent—Global ocean modeling with GPU acceleration in python}",
  author={H{\"a}fner, Dion and Nuterman, Roman and Jochum, Markus},
  journal={Journal of Advances in Modeling Earth Systems},
  volume={13},
  number={12},
  pages={e2021MS002717},
  year={2021},
  publisher={Wiley Online Library}
}

@article{ramadhan2020oceananigans,
  title="{Oceananigans. jl: Fast and friendly geophysical fluid dynamics on GPUs}",
  author={Ramadhan, Ali and Wagner, Gregory and Hill, Chris and Campin, Jean-Michel and Churavy, Valentin and Besard, Tim and Souza, Andre and Edelman, Alan and Ferrari, Raffaele and Marshall, John},
  journal={Journal of Open Source Software},
  volume={5},
  number={53},
  year={2020}
}

@article{rasp2018deep,
  title="{Deep learning to represent subgrid processes in climate models}",
  author={Rasp, Stephan and Pritchard, Michael S and Gentine, Pierre},
  journal={Proceedings of the national academy of sciences},
  volume={115},
  number={39},
  pages={9684--9689},
  year={2018},
  publisher={National Academy of Sciences}
}

@article{kochkov2024neural,
  title="{Neural general circulation models for weather and climate}",
  author={Kochkov, Dmitrii and Yuval, Janni and Langmore, Ian and Norgaard, Peter and Smith, Jamie and Mooers, Griffin and Kl{\"o}wer, Milan and Lottes, James and Rasp, Stephan and D{\"u}ben, Peter and others},
  journal={Nature},
  volume={632},
  number={8027},
  pages={1060--1066},
  year={2024},
  publisher={Nature Publishing Group UK, London}
}

@article{lam2023learning,
  title="{Learning skillful medium-range global weather forecasting}",
  author={Lam, Remi and Sanchez-Gonzalez, Alvaro and Willson, Matthew and Wirnsberger, Peter and Fortunato, Meire and Alet, Ferran and Ravuri, Suman and Ewalds, Timo and Eaton-Rosen, Zach and Hu, Weihua and others},
  journal={Science},
  volume={382},
  number={6677},
  pages={1416--1421},
  year={2023},
  publisher={American Association for the Advancement of Science}
}

@article{fuhrer2018near,
  title="{Near-global climate simulation at 1 km resolution: establishing a performance baseline on 4888 GPUs with COSMO 5.0}",
  author={Fuhrer, Oliver and Chadha, Tarun and Hoefler, Torsten and Kwasniewski, Grzegorz and Lapillonne, Xavier and Leutwyler, David and L{\"u}thi, Daniel and Osuna, Carlos and Sch{\"a}r, Christoph and Schulthess, Thomas C and others},
  journal={Geoscientific Model Development},
  volume={11},
  number={4},
  pages={1665--1681},
  year={2018},
  publisher={Copernicus Publications, G{\"o}ttingen, Germany}
}

@article{kay2015community,
  title="{The Community Earth System Model (CESM) large ensemble project: A community resource for studying climate change in the presence of internal climate variability}",
  author={Kay, Jennifer E and Deser, Clara and Phillips, A and Mai, Angeline and Hannay, Cecile and Strand, Gary and Arblaster, Julie Michelle and Bates, Susan C and Danabasoglu, Gokhan and Edwards, James and others},
  journal={Bulletin of the American Meteorological Society},
  volume={96},
  number={8},
  pages={1333--1349},
  year={2015}
}

@article{williamson1992standard,
  title={A standard test set for numerical approximations to the shallow water equations in spherical geometry},
  author={Williamson, David L and Drake, John B and Hack, James J and Jakob, R{\"u}diger and Swarztrauber, Paul N},
  journal={Journal of computational physics},
  volume={102},
  number={1},
  pages={211--224},
  year={1992},
  publisher={Elsevier}
}

@inproceedings{li2021fourier,
  author    = {Li, Zongyi and Kovachki, Nikola and
               Azizzadenesheli, Kamyar and Liu, Burigede and
               Bhattacharya, Kaushik and Stuart, Andrew and
               Anandkumar, Anima},
  title     = "{{Fourier} Neural Operator for Parametric
               Partial Differential Equations}",
  booktitle = {International Conference on Learning Representations},
  year      = {2021},
  url       = {https://arxiv.org/abs/2010.08895}
}

@inproceedings{chen2018neural,
  author    = {Chen, Ricky T. Q. and Rubanova, Yulia and
               Bettencourt, Jesse and Duvenaud, David},
  title     = "{Neural Ordinary Differential Equations}",
  booktitle = {Advances in Neural Information Processing Systems},
  volume    = {31},
  year      = {2018},
  url       = {https://arxiv.org/abs/1806.07366}
}

@article{raissi2019physics,
  author  = {Raissi, Maziar and Perdikaris, Paris and
             Karniadakis, George Em},
  title   = "{Physics-informed neural networks: A deep learning
             framework for solving forward and inverse problems
             involving nonlinear partial differential equations}",
  journal = {Journal of Computational Physics},
  volume  = {378},
  pages   = {686--707},
  year    = {2019},
  doi     = {10.1016/j.jcp.2018.10.045}
}

@inproceedings{trautner2020informative,
  title="{Informative ensemble Kalman learning for neural structure}",
  author={Trautner, Margaret and Margolis, Gabriel and Ravela, Sai},
  booktitle={International Conference on Dynamic Data Driven Application Systems},
  pages={191--199},
  year={2020},
  organization={Springer}
}

@article{trautner2021learn,
  title="{Learn like the pro: Norms from theory to size neural computation}",
  author={Trautner, Margaret and Li, Ziwei and Ravela, Sai},
  journal={arXiv preprint arXiv:2106.11409},
  year={2021}
}

@article{peixoto2016accuracy,
  title="{Accuracy analysis of mimetic finite volume operators on geodesic grids and a consistent alternative}",
  author={Peixoto, Pedro S},
  journal={Journal of Computational Physics},
  volume={310},
  pages={127--160},
  year={2016},
  publisher={Elsevier}
}

@article{calandrini2021comparing,
  title="{Comparing numerical accuracy and stability for different horizontal discretizations in MPAS-Ocean}",
  author={Calandrini, Sara and Engwirda, Darren and Petersen, Mark},
  journal={Ocean Modelling},
  volume={168},
  pages={101908},
  year={2021},
  publisher={Elsevier}
}

@article{yu2020comparing,
  title="{Comparing numerical accuracy of icosahedral A-grid and C-grid schemes in solving the shallow-water model}",
  author={Yu, Yonggang G and Wang, Ning and Middlecoff, Jacques and Peixoto, Pedro S and Govett, Mark W},
  journal={Monthly Weather Review},
  volume={148},
  number={10},
  pages={4009--4033},
  year={2020}
}

@article{linck2026evaluating,
  title="{Evaluating LLMs for Translating Legacy Fortran Atmospheric Physics to GPU-Accelerated JAX}",
  author={Linck, Iris and Dennis, John and Schreck, John and Stengel, Karen and Gagne II, David John},
  journal={arXiv preprint arXiv:2610.12662},
  year={2026}
}

@article{yuan2026wrf,
  title="{WRF-JAX v1. 1.28: a differentiable regional weather model with fully coupled physics, numerically consistent with WRF and matching 128-process CPU performance on a single GPU}",
  author={Yuan, Shijin and Shen, Tianyu and Mu, Bin and Wei, Zhen and Qin, Bo and Dai, Guokun},
  journal={arXiv preprint arXiv:2610.35095},
  year={2026}
}

@article{gentine2026legoesm,
  title="{legoESM: a modular, differentiable, multiscale, AI-ready Earth system model built with AI agents}",
  author={Gentine, Pierre and Balwada, Dhruv and Pa{\c{c}}al, Ayta{\c{c}} and Hawkins, Linnia and Adcroft, Alistair and Fan, Hang and Fang, Jianing and Lahlou, Aya and Lamb, Kara D and Lawrence, David M and others},
  journal={arXiv preprint arXiv:2610.11883},
  year={2026}
}

@article{feng2024,
  author  = {Feng, Dongyu and Tan, Zeli and Engwirda, Darren and Wolfe, Jonathan D. and Xu, Donghui and Liao, Chang and Bisht, Gautam and Benedict, James J. and Zhou, Tian and Li, Hong-Yi and Leung, L. Ruby},
  title   = {Simulation of compound flooding using river-ocean two-way coupled {E3SM} ensemble on variable-resolution meshes},
  journal = {Journal of Advances in Modeling Earth Systems},
  volume  = {16},
  number  = {6},
  pages   = {e2023MS004054},
  year    = {2024},
  doi     = {10.1029/2023MS004054}
}

@article{lilly2025,
author  = {Lilly, Jeremy R. and Capodaglio, Giacomo and Engwirda, Darren and Higdon, Robert L. and Petersen, Mark R.},
title   = {Local time-stepping for the shallow water equations using {CFL} optimized forward-backward Runge-Kutta schemes},
journal = {Journal of Computational Physics},
volume  = {520},
pages   = {113511},
year    = {2025},
doi     = {10.1016/j.jcp.2024.113511}
}

@article{lilly2023,
author  = {Lilly, Jeremy R. and Capodaglio, Giacomo and Petersen, Mark R. and Brus, Steven R. and Engwirda, Darren and Higdon, Robert L.},
title   = {Storm Surge Modeling as an Application of Local Time-Stepping in {MPAS-Ocean}},
journal = {Journal of Advances in Modeling Earth Systems},
year    = {2023},
volume  = {15},
number  = {1},
pages   = {e2022MS003327},
doi     = {10.1029/2022MS003327}
}

@article{brus2026,
author  = {Brus, S. R. and Wirasaet, D. and Engwirda, D. and Blakely, C. P. and Titterton, M. and Petersen, M. R. and Begeman, C. B. and Kennedy, A. B. and Westerink, J. J.},
title   = {A Subgrid Correction Scheme for Accurate km-Scale Coastal Inundation in the Energy Exascale Earth System Model},
journal = {Journal of Advances in Modeling Earth Systems},
year    = {2026},
volume  = {18},
number  = {8},
pages   = {e2026MS005842},
doi     = {10.1029/2026MS005842}
}

@misc{trautner2019,
  author        = {Margaret Trautner and Sai Ravela},
  title         = {Neural Integration of Continuous Dynamics},
  year          = {2019},
  eprint        = {1911.10309},
  archivePrefix = {arXiv},
  url           = {https://arxiv.org/abs/1911.10309}
}

@misc{todorovic2026,
  author        = {Luka Todorovic and Kaushal Gianchandani
                   and Sai Ravela},
  title         = {The Physics Is the Network:
                   Fast, Training-Free Neural Emulation},
  year          = {2026},
  eprint        = {2610.20525},
  archivePrefix = {arXiv},
  url           = {https://arxiv.org/abs/2610.20525}
}

@article{thuburn2009,
  author  = {Thuburn, John and Ringler, Todd D. and
             Skamarock, William C. and Klemp, Joseph B.},
  title   = {Numerical representation of geostrophic modes
             on arbitrarily structured {C}-grids},
  journal = {Journal of Computational Physics},
  year    = {2009},
  volume  = {228},
  pages   = {8321--8335},
  doi     = {10.1016/j.jcp.2009.08.006}
}

@article{ringler2010,
  author  = {Ringler, Todd D. and Thuburn, John and
             Klemp, Joseph B. and Skamarock, William C.},
  title   = {A unified approach to energy conservation and
             potential vorticity dynamics for
             arbitrarily-structured {C}-grids},
  journal = {Journal of Computational Physics},
  year    = {2010},
  volume  = {229},
  pages   = {3065--3090},
  doi     = {10.1016/j.jcp.2009.12.007}
}

@article{chen2011,
  author  = {Chen, Qiang and Gunzburger, Max and Ringler, Todd},
  title   = {A Scale-Invariant Formulation of the
             Anticipated Potential Vorticity Method},
  journal = {Monthly Weather Review},
  year    = {2011},
  volume  = {139},
  pages   = {2614--2629},
  doi     = {10.1175/MWR-D-10-05004.1}
}

@misc{mpas_model,
  author       = {{MPAS-Model developers}},
  title        = {{MPAS-Model}, version 8.4.2},
  howpublished = {\url{https://github.com/MPAS-Dev/MPAS-Model/tree/v8.4.2}},
}

@misc{jax2018github,
  author = {James Bradbury and Roy Frostig and Peter Hawkins and
            Matthew James Johnson and Yash Katariya and Chris Leary and
            Dougal Maclaurin and George Necula and Adam Paszke and
            Jake Vander{P}las and Skye Wanderman-{M}ilne and Qiao Zhang},
  title  = {{JAX}: composable transformations of
            {Python}+{NumPy} programs},
  year   = {2018},
  url    = {https://github.com/jax-ml/jax}
}

@misc{rackauckas2020ude,
  author = {Christopher Rackauckas and Yingbo Ma and Julius Martensen and Collin Warner and Kirill Zubov and Rohit Supekar and Dominic Skinner and Ali Ramadhan and Alan Edelman},
  title = {Universal Differential Equations for Scientific Machine Learning},
  year = {2020},
  note = {arXiv:2001.04385, first submitted January 13, 2020},
  url = {https://arxiv.org/abs/2001.04385}
}

@article{lopezgomez2022ensemble,
  author = {Ignacio Lopez-Gomez and Costa Christopoulos and Haakon Ludvig Langeland Ervik and Oliver R. A. Dunbar and Yair Cohen and Tapio Schneider},
  title = {Training Physics-Based Machine-Learning Parameterizations With Gradient-Free Ensemble Kalman Methods},
  journal = {Journal of Advances in Modeling Earth Systems},
  volume = {14},
  number = {8},
  pages = {e2022MS003105},
  year = {2022},
  doi = {10.1029/2022MS003105},
  url = {https://doi.org/10.1029/2022MS003105}
}

@misc{ravela2026enscgp,
  author = {Sai Ravela and Jae Deok Kim and Kenneth Gee and Xingjian Yan and Samson Mercier and Lubna Albarghouty and Anamitra Saha},
  title = {Ensemble-Conditional Gaussian Processes (Ens-CGP): Representation, Geometry, and Inference},
  year = {2026},
  note = {arXiv:2602.13871},
  url = {https://arxiv.org/abs/2602.13871}
}

@misc{gee2026cyclone,
  author = {Kenneth Gee and Sai Ravela},
  title = {Learning a Stochastic Differential Equation Model of Tropical Cyclone Intensification from Reanalysis and Observational Data},
  year = {2026},
  note = {arXiv:2601.08116, version 3},
  url = {https://arxiv.org/abs/2601.08116}
}

@article{ravela2007smoothing,
  author = {Sai Ravela and Dennis McLaughlin},
  title = {Fast Ensemble Smoothing},
  journal = {Ocean Dynamics},
  year = {2007},
  volume = {57},
  number = {2},
  pages = {123--134},
  doi = {10.1007/s10236-006-0098-6}
}

@article{ravela2010observatory,
  author = {Sai Ravela and John Marshall and Chris Hill and Andrew Wong and Scott Stransky},
  title = {A Realtime Observatory for Laboratory Simulation of Planetary Flows},
  journal = {Experiments in Fluids},
  year = {2010},
  volume = {48},
  number = {5},
  pages = {915--925},
  doi = {10.1007/s00348-009-0752-0}
}

@article{saha2024rainfall,
  author = {Anamitra Saha and Sai Ravela},
  title = {Statistical-Physical Adversarial Learning From Data and Models for Downscaling Rainfall Extremes},
  journal = {Journal of Advances in Modeling Earth Systems},
  year = {2024},
  volume = {16},
  number = {6},
  pages = {e2023MS003860},
  doi = {10.1029/2023MS003860}
}

@article{kim2026npi,
  author = {Jae Deok Kim and Sai Ravela and Rob L. Evans},
  title = {The Neuro-Physical Inverter: A Modular Framework for Magnetotelluric Inversion Coupling Ensemble Conditioning with Residual Learning},
  journal = {IEEE Transactions on Geoscience and Remote Sensing},
  year = {2026},
  doi = {10.1109/TGRS.2026.3740359},
  note = {Accepted for publication; arXiv:2610.03225v2}
}
\bibliographystyle{sciencemag}



\section*{Acknowledgments}
We acknowledge the MPAS-Model development community for openly sharing its shallow-water and ocean implementations, which informed selected numerical treatments in \phispin\ and provided reference models for evaluation. We also acknowledge the developers and contributors of JAX and the scientific Python ecosystem on which our implementation builds. Specific adaptations and applicable software notices are documented in the code repository. ChatGPT (OpenAI) and Grammarly were used to assist with drafting, revising, and editing the manuscript. The authors retain responsibility for the content, accuracy, and integrity of the work, including the ideas, analyses, interpretations, and conclusions.

\paragraph*{Funding:} This material is based in part upon work supported by the Bangkok Bank, Community Jameel for Jameel Observatory CREWSnet, MIT MSRP program, and by the MIT Climate Grand Challenges Weather and Climate Extremes projects support from Schmidt Sciences. K.G. was supported by an appointment to the NASA Postdoctoral Program at NASA Goddard Institute for Space Studies from the NASA Astrobiology Program, administered by Oak Ridge Associated Universities, under contract with NASA.


\paragraph*{Competing interests:}
There are no competing interests to declare.

\paragraph*{Data, code and materials availability:}
The \phispinsw\ source code and the configurations, data and analysis scripts supporting this preprint will be made available upon publication of a peer-reviewed article.

\subsection*{Supplementary materials}
Methods\\
Supplementary Text\\
Figs. S1 to S4\\


\newpage


\renewcommand{\thefigure}{S\arabic{figure}}
\renewcommand{\thetable}{S\arabic{table}}
\renewcommand{\theequation}{S\arabic{equation}}
\renewcommand{\thepage}{S\arabic{page}}
\setcounter{figure}{0}
\setcounter{table}{0}
\setcounter{equation}{0}
\setcounter{page}{1} 


\begin{center}
\section*{Supplementary Materials for\\ \scititle}


Kaushal Gianchandani$^{\ast}$,
Luka Todorovic
and
Sai Ravela$^{\dagger}$\\
$^{\ast,\dagger}$Corresponding authors. Emails: kaushalg@mit.edu ($\ast$), ravela@mit.edu ($\dagger$)\\
\end{center}

\subsubsection*{This PDF file includes:}
Methods\\
Supplementary Text\\
Figures S1 to S4\\

\newpage


\section*{Methods}
\subsection*{Numerical implementation of the shallow-water equations}
\phispinsw\ evaluates the governing equations presented in the main text using a compatible C-grid discretisation on spherical Voronoi grids. The spatial operators follow Thuburn et al.~\cite{thuburn2009} and Ringler et al.~\cite{ringler2010}, with implementation details informed by the openly available MPAS shallow-water model \cite{mpas_model}.

The grid comprises Voronoi cells and their dual Delaunay triangulation. Layer thickness and bottom elevation are stored at cell centers, prognostic normal velocity at edges, and relative vorticity and potential vorticity at Voronoi vertices. Grid-point positions are represented by Cartesian unit vectors. Cell areas, edge lengths and distances between neighboring cell centers supply the geometric factors in the discrete operators. The spatial operators and time integration are implemented in JAX \cite{jax2018github}, using float64 arithmetic.

The following expressions describe the unforced, all-active core. For cell $i$, let $A_i$ denote its area. Edge $e$ has length $l_e$, dual length $d_e$, and a normal orientation from its left cell $L$ to its right cell $R$. The compatible divergence and gradient are:
\begin{equation}
(\mathcal{D}F)_i
= \frac{1}{A_i}\sum_{e\in\partial i} n_{ie}l_eF_e,
\quad
(\mathcal{G}\psi)_e = \frac{\psi_R-\psi_L}{d_e},
\end{equation}
where $n_{ie}=+1$ for an outward-oriented edge and $-1$ for an inward-oriented edge. Edge thickness is the arithmetic mean of the adjacent cell values, $\widehat{h}_e=(h_L+h_R)/2$. The shared mass flux and thickness tendency are:
\begin{equation}
F_e=\widehat{h}_eu_e,
\quad
\dot{h}_i=-(\mathcal{D}F)_i.
\end{equation}

Relative vorticity is evaluated from the signed circulation around each dual cell. Dual-cell thickness is obtained using primal--dual intersection-area weights, and vertex potential vorticity is diagnosed by exact division, $q_v=(\zeta_v+f_v)/h_v$. Cell kinetic energy is
\begin{equation}
K_i=\frac{1}{A_i}\sum_{e\in\partial i}\frac{l_ed_e}{4}u_e^2.
\end{equation}
The momentum tendency combines the energy-neutral PV-flux mapping of Ringler et al.~\cite{ringler2010} with the gradient of the Bernoulli potential, $B_i=K_i+g(h_i+b_i)$. The implementation evaluates $-\mathcal{G}(gh)$, $-\mathcal{G}K$ and $-\mathcal{G}(gb)$ separately before combining their contributions. The PV mapping uses symmetric averaging of edge PV together with antisymmetric geometric coupling weights. The reconstruction weights follow
Thuburn et al.~\cite{thuburn2009}.

A single flux is shared by the cells adjacent to each interior edge, giving equal and opposite contributions to their mass budgets. The energy-neutral property concerns the spatial PV-flux operator; it does not imply exact energy conservation by the fully discrete model.

The discrete tendency is implemented as an explicit network of callable array operators connected through named intermediate fields. Each operator acts on fields at their native cell, edge or vertex locations. Linear interpolation and differential operators are combined with products, quadratic kinetic-energy evaluation, exact PV division and additions of tendency contributions. This network is the executed tendency calculation, rather than a diagram constructed independently of the solver. The same operation definitions support numerical evaluation and on-demand inspection of the computational dependencies. The complete discrete-time evolution additionally includes the recurrent RK4 stage calculations. This directly coded computation is functionally the prescribed neural-integration graph, even though a separate graph of individual PolyNet nodes is not materialized \cite{trautner2019}.

Within this equation-derived PolyNet-like representation, geometry-dependent stencil coefficients and prescribed physical and numerical parameters provide the weights; they are not inferred from training data. Exact PV division is retained, so the unforced core, without state-dependent protection, is rational rather than exclusively polynomial. Configured forcing, friction and masking operations also remain explicit and are not replaced by polynomial approximations. No training data or learned closure is required to advance the state.

For domains with a fixed wet mask, activity masks enter state preparation, transport, PV diagnosis and tendency assembly. Inactive vertices use a safe auxiliary denominator before their PV is masked to zero; division at active vertices is unchanged. Active--inactive interfaces have zero normal transport.
No tangential no-slip condition is imposed. These fixed masks do not represent moving shorelines or coastal inundation. Optional state-dependent drying protection is applied separately through stage-wise transport and momentum projections, preserving the distinction between prognostic velocity and the velocity used to construct mass flux. Configured thickness and momentum sources are evaluated separately and added to the corresponding tendencies.

Connectivity and geometric coefficients are prepared before integration. Each spatial operator gathers neighbouring field values and combines them using the prescribed weights, without forming dense global operator matrices. The evolving fields remain JAX arrays, and just-in-time compilation produces an executable for the selected CPU or GPU. Operator-level nodes need not correspond one-to-one to device kernels, and the inspection graph is not rebuilt at each timestep. Automatic differentiation is supported through verified dynamical paths with explicit physical parameters, but is not assumed through grid generation, file input or switching thresholds. Such differentiation is optional: it is not used to train the present fixed-physics model, and the intended ensemble-conditioned learning mechanism does not require it. The semantic reconstruction and adaptive extension of the full \phispin\ architecture are not implemented or benchmarked in the present version.

The baseline solver advances the complete tendency using classical four-stage, fourth-order Runge--Kutta integration. State-dependent terms are recomputed at each provisional state, and time-dependent providers receive the physical stage times $t$, $t+\Delta t/2$, $t+\Delta t/2$ and $t+\Delta t$. The four stage evaluations are unrolled during tracing; repeated timesteps are executed using \texttt{jax.lax.scan}. The Runge--Kutta coefficients are fixed and non-trainable. Fourth-order temporal accuracy applies to the underlying smooth RK4 formulation; it is not asserted when state-dependent drying projections activate.

Optional anticipated potential-vorticity stabilisation follows the APVM approach \cite{chen2011}, using the timestep-based formulation and PV reconstruction documented in MPAS-SW \cite{mpas_model}. When enabled, APVM uses the full model timestep at each Runge--Kutta stage to modify the edge PV entering the rotational momentum tendency. For each experiment, we specify whether APVM is enabled and the coefficient used.

\subsection*{AI-assisted software development}
We used OpenAI Codex as a coding assistant to develop \phispinsw, including source-code inspection, JAX implementation, debugging, testing and documentation. We specified the physical formulation, numerical choices and verification criteria, using published C-grid methods and pinned MPAS source code as numerical references. Analytical benchmarks and comparisons with MPAS assessed the resulting solver.

\subsection*{Numerical simulations}
We evaluated analytical accuracy, agreement with established numerical models and computational cost in separate experiments. All simulations used double-precision floating-point arithmetic (IEEE 754 binary64). Gravity was $g=9.80616~m~s^{-2}$ throughout. The spherical radius was $a=6371220~m$ and the rotation rate was set to $\Omega=7.292\times10^{-5}~s^{-1}$ for all simulations.

The global icosahedral grids were labelled G4--G9 and contained 2562, 10242, 40962, 163842, 655362 and 2621442 cells, respectively. Their typical horizontal spacings were approximately $446$, $223$, $112$, $56$, $28$ and $14~km$. We defined typical spacing as the square root of the mean cell area.

\paragraph{Analytical benchmarks:} We used Williamson tests 1-4 \cite{williamson1992standard} to assess prescribed transport, the maintenance of balanced flows and the response to time-dependent forcing. These experiments used closed spherical domains with APVM and physical damping disabled.

In test 1, a prescribed solid-body flow transported a $1000~m$ cosine bell of radius $a/3$ around the equator over $12~days$. We added a uniform $1000~m$ pedestal to maintain positive layer thickness. Tests 2 and 3 examined the aligned global and compact-support balanced flows, respectively, using the speed parameter $2\pi a/(12~days)$ and reference geopotential $29400~m^{2}~s^{-2}$. We integrated test 2 for $15~days$ on G4--G7 grids and evaluated refinement errors at day 5. Test 3 ran for $5~days$ on G6. Test 4 followed the forced translating low for $5~days$ with a background jet parameter of $20~m~s^{-1}$; we evaluated its analytical mass and momentum sources at the RK4 stage times. Tests 1 and 4 also used G6. For each experiment, we selected a fixed timestep from the initial state using a CFL target of 0.4 and a $600~s$ cap, and saved scientific output every $6~hours$.

We compared the simulated fields with the analytical solutions at the model's cell centres and saved output times. We reported the mean absolute error ($L_1$), root-mean-square error ($L_2$) and maximum absolute error ($L_\infty$). Cells contributed to the averages in proportion to their area. To obtain relative errors, we divided each measure by the same measure of the analytical reference field. For height errors, the reference was the bell above its uniform pedestal in test 1, the low relative to its background height in test 4, and the full height field in tests 2 and 3. We used a least-squares fit to reconstruct cell-centre velocities from edge-normal velocities, then compared the resulting vectors with the analytical velocities. For test 4, our relative velocity errors used the full flow, including the background jet; the original test definition removed that jet from the reference. We ran test 2 at four grid resolutions (G4--G7) to examine how its error changed as the grid became finer. Tests 1, 3 and 4 were run only at G6.

\paragraph{Mountain-flow comparison:} We used Williamson test 5 to examine agreement with MPAS-SW during extended nonlinear evolution. The experiment ran for $50~days$ on G6 with a $150~s$ timestep. We initialized a balanced flow with a $5960~m$ reference surface and $20~m~s^{-1}$ equatorial velocity over a $2000~m$ conical mountain centred at $270^\circ$ E, $30^\circ$ N, with the prescribed $20^\circ$ longitude--latitude radius. APVM and explicit damping remained disabled.

We exported the native grid, topography and day-zero state to MPAS-SW, verifying entity correspondence, edge orientation and initialized fields. MPAS regenerated its geometric coefficients, and both models advanced independently using RK4. We compared height at day 15, PV at day 50 and native mass, energy and potential-enstrophy histories relative to each model's initial values. Differences were \phispinsw\ minus MPAS-SW. The reference used pinned MPAS-SW v8.4.2.

\paragraph{Computational benchmarks:} We measured computational cost in separate Williamson-5 trials on MPAS-derived grids and initial states at G5--G9. Timesteps were $300$, $150$, $75$, $30$ and $15~s$, respectively. APVM was disabled in both models. We performed ten separate 2000-step trials per configuration, restarting each from the same initial state and retaining the first GPU execution after compilation. \phispinsw\ ran on individual NVIDIA A100, H200 and B200 GPUs. MPAS-SW ran with 32, 64 and 96 MPI ranks on an AMD EPYC 9654 package, with one distinct physical core and one computational thread per rank. We reported all three rank counts separately.

We excluded initialization, compilation and file output from the timings. GPU timings included RK4 and its state checks, and ended only after the GPU had finished. For MPAS-SW, we used the longest total integration time reported by any MPI process, including tracer calculations and communication between processes. We calculated steps per second and steps per joule for each trial, then reported their medians and minimum–maximum ranges. Energy measurements covered the CPU package or GPU board. They included idle energy but excluded the computer hosting the GPU. The reported ranges showed variation among trials, not confidence intervals. These comparisons described the measured components, not the energy use of the whole computer.

\paragraph{Idealized coastal cyclone:} We examined coastal response in a four-day idealized Gulf of Thailand experiment, comparing \phispinsw\ with one-layer MPAS-Ocean. Both models began from rest with zero surface elevation and used common bathymetry and prescribed geographic wind-stress samples in Pa. The shared wet domain contained 58328 cells, and both models advanced with a $2~s$ timestep. We used a reference density of $1026~kg~m^{-3}$ and a quadratic bottom-drag coefficient of 0.0025. APVM and tides were disabled, while drying protection retained a $1~m$ residual thickness.

\phispinsw\ retained the full mesh with a fixed wet mask and impermeable free-slip coasts; MPAS-Ocean used the corresponding culled wet mesh and its native coastal operators. We supplied stress directly, without conversion from wind speed or a landfall-triggered drag switch. \phispinsw\ interpolated forcing samples linearly at stage times, whereas MPAS-Ocean held scheduled samples.

We compared 577 outputs at $600~s$ intervals on corresponding native cells, without intermodel interpolation, and referenced time to prescribed landfall at elapsed day 2. We evaluated surface elevation, reconstructed speed and water volume. Coastal histories sampled the nearest modeled coastal cells to $(99.2^\circ\mathrm{E},9.5^\circ\mathrm{N})$, $(99.8^\circ\mathrm{E},11.5^\circ\mathrm{N})$ and $(100.5^\circ\mathrm{E},13.0^\circ\mathrm{N})$ as illustrative model locations. We weighted signed \phispinsw\ minus MPAS-Ocean differences by wet-cell area. Empirical 2.5th--97.5th percentiles described the central 95\% of their distributions, not confidence intervals. Configurations, source identities, input checksums and analysis selections were retained.

\section*{Supplementary Text}
\begin{figure}[!ht]
\centering
\includegraphics[width=\textwidth]{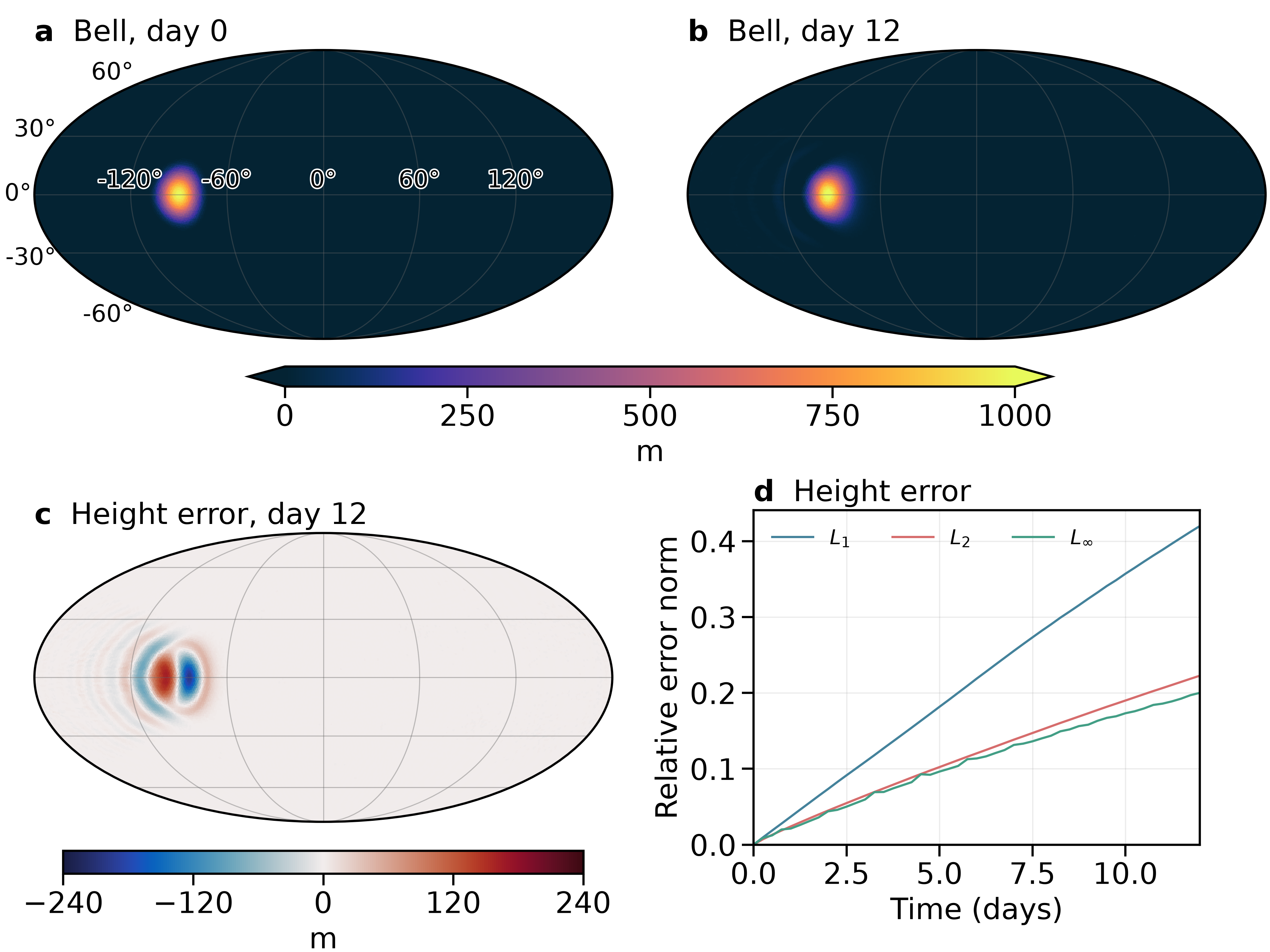}
\caption{
\textbf{Williamson test 1: cosine-bell transport.}
Bell height above the uniform $1000~m$ pedestal at day 0 (a)
and day 12 (b), signed height error at day 12 (c), and relative
$L_1$, $L_2$ and $L_\infty$ error histories (d).
Panels a and b share the same colour scale.
Errors are \phispinsw\ minus the analytical solution,
with each relative norm normalized by the corresponding
analytical bell-anomaly norm. The simulation uses the G6 grid.
}
\label{fig:w1}
\end{figure}

Williamson test 1 isolates prescribed transport by carrying a cosine bell around the sphere in 12 days (Supplementary Fig.~\ref{fig:w1}). On G6, the relative $L_2$ and $L_\infty$ height errors reach approximately 0.22 and 0.20 after one circuit. Both use the analytical bell anomaly for normalization.

\begin{figure}[!ht]
\centering
\includegraphics[width=\textwidth]{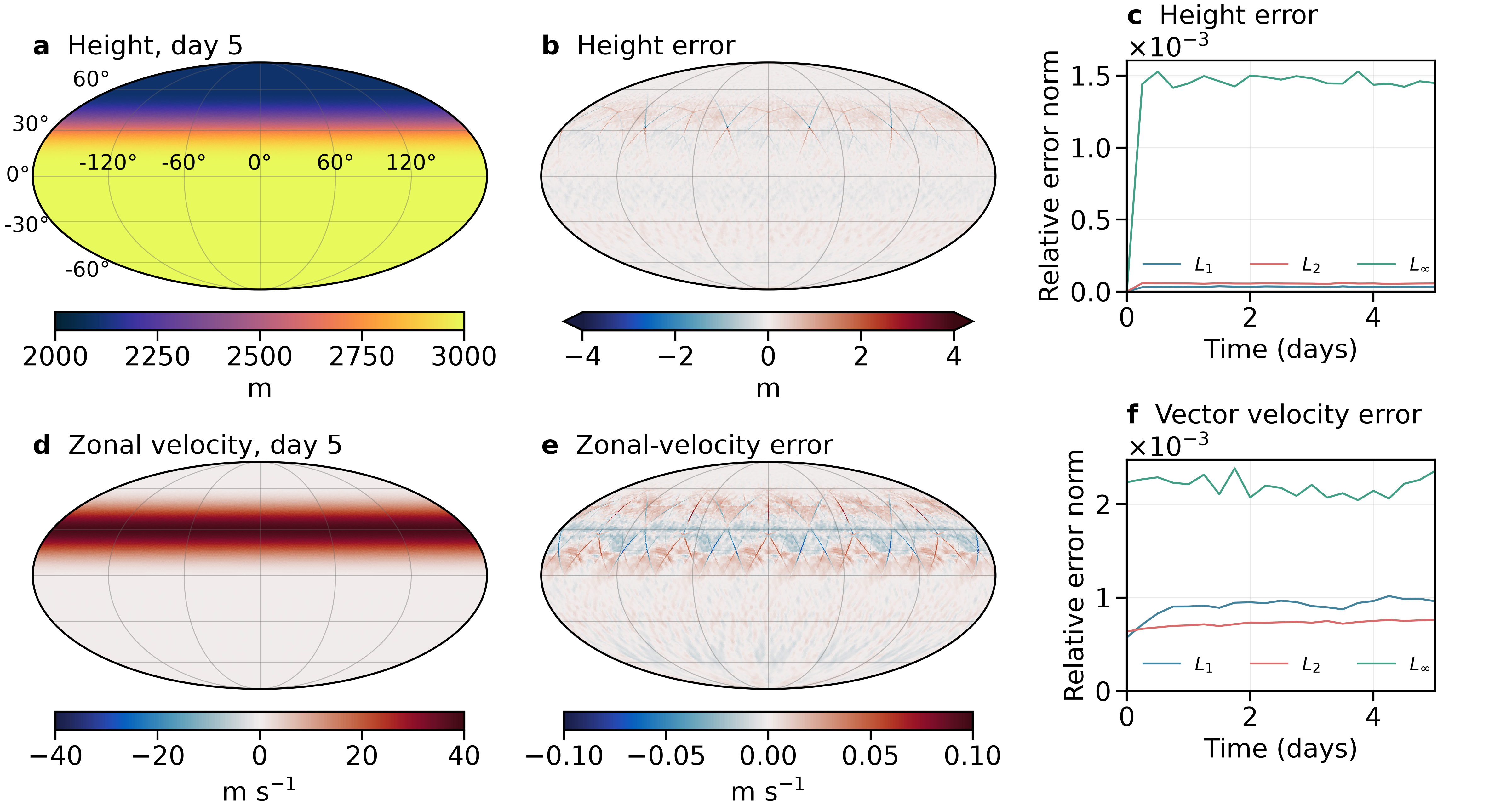}
\caption{
\textbf{Williamson test 3: steady balanced jet.}
Native-grid height (a) and reconstructed zonal velocity (d)
at day 5, their signed errors relative to the analytical
solution (b,e), and relative $L_1$, $L_2$ and $L_\infty$
error histories for height (c) and vector velocity (f).
Errors are \phispinsw\ minus the analytical solution;
each relative norm uses the corresponding analytical norm.
The simulation uses the G6 grid.
}
\label{fig:w3}
\end{figure}

Williamson test 3 examines a steady balanced jet confined to a limited latitude range (Supplementary Fig.~\ref{fig:w3}). At day 5 on G6, the relative $L_2$ errors are $5.6\times10^{-5}$ for height and $7.6\times10^{-4}$ for vector velocity.

\begin{figure}[!ht]
\centering
\includegraphics[width=\textwidth]{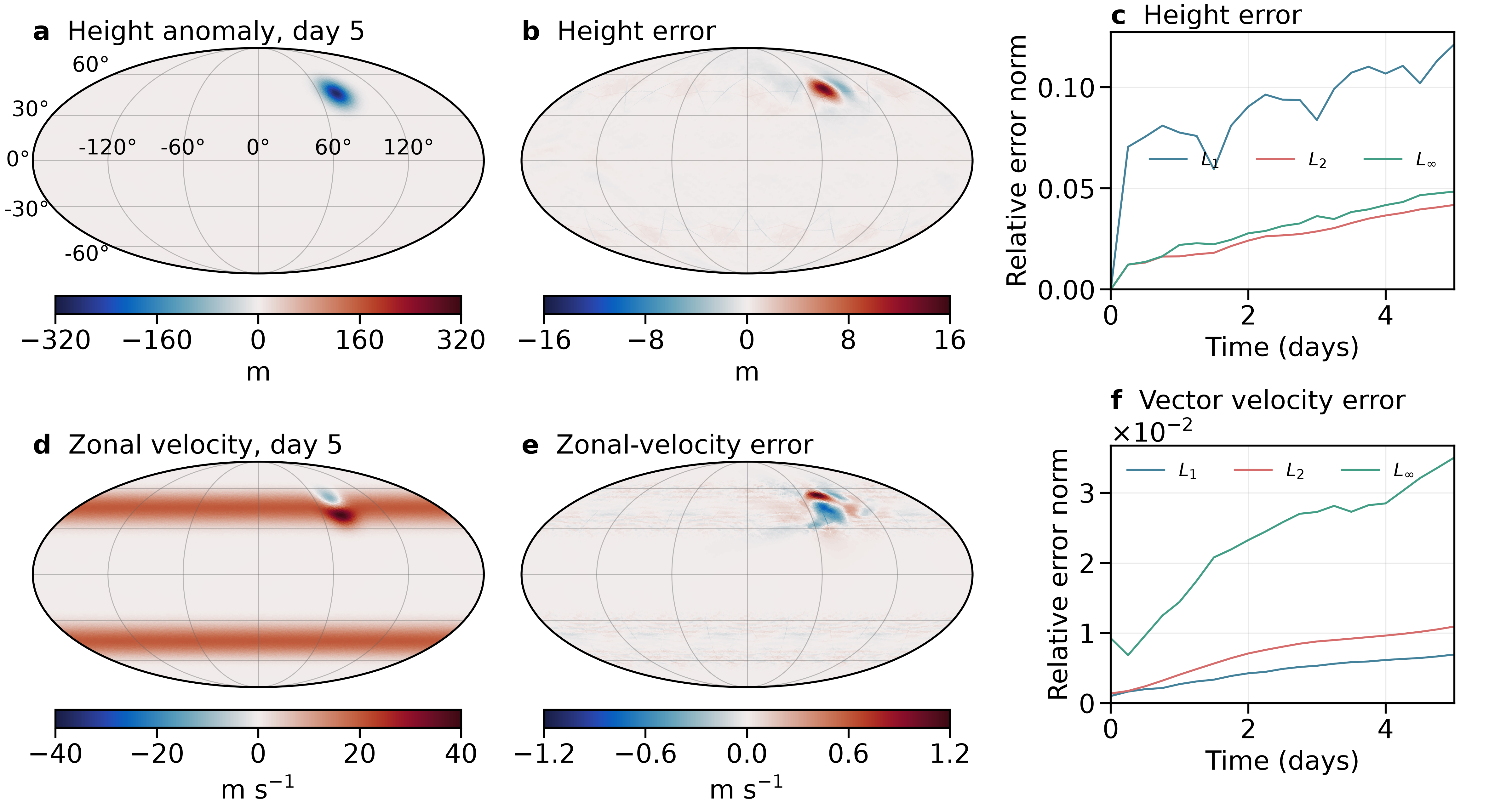}
\caption{
\textbf{Williamson test 4: forced translating low.}
Native-grid height anomaly relative to the prescribed background
(a) and reconstructed zonal velocity (d) at day 5, their signed
errors relative to the analytical solution (b,e), and relative
$L_1$, $L_2$ and $L_\infty$ error histories for height (c)
and vector velocity (f). Errors are \phispinsw\ minus
the analytical solution. Height norms are normalized by the
analytical low anomaly; velocity norms use the full analytical
flow, including the background jet. The simulation uses the G6 grid.
}
\label{fig:w4}
\end{figure}

Williamson test 4 examines a translating low maintained by prescribed time-dependent forcing (Supplementary Fig.~\ref{fig:w4}). At day 5 on G6, the relative $L_2$ errors are approximately 0.042 for the height anomaly and 0.011 for vector velocity. Height errors use the analytical low anomaly for normalization, whereas velocity errors use the full analytical flow, including the background jet. These single-resolution tests do not establish convergence.


\begin{figure}[!ht]
\centering
\includegraphics[width=\textwidth]{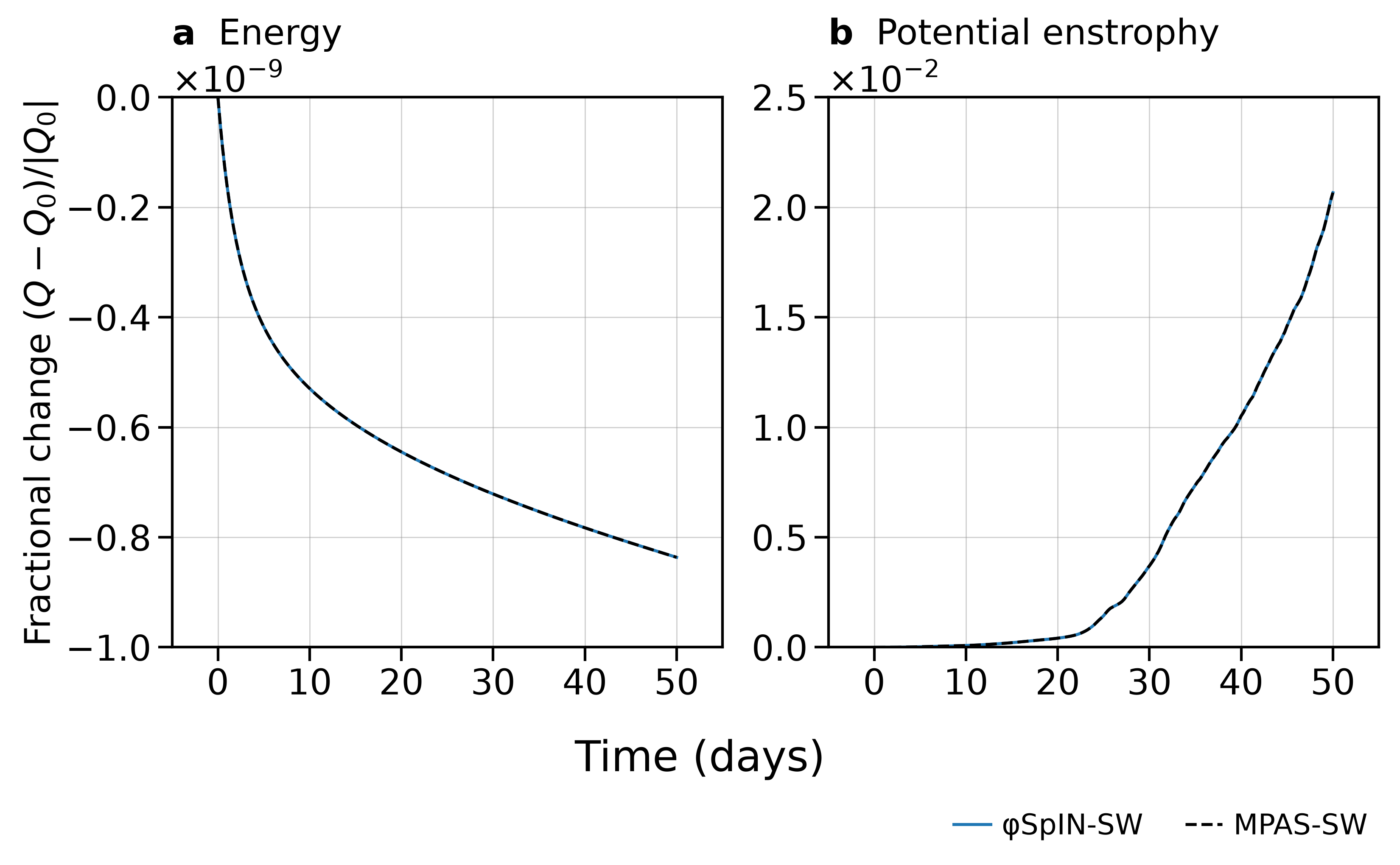}
\caption{
\textbf{Williamson test 5: conservation diagnostics.} Fractional changes $(Q-Q_0)/|Q_0|$ in native-grid total energy (a) and potential enstrophy (b) over 50 days, where $Q_0$ is each model's initial value. Solid blue and dashed black lines show \phispinsw\ and MPAS-SW, respectively. Both simulations use the same G6 grid (40,962 cells, typical resolution $\sim112~km$) and a $150~s$ timestep.
}
\label{fig:w5-conservation}
\end{figure}

\end{document}